\documentclass[aps,prb,twocolumn,superscriptaddress,longbibliography,footinbib]{revtex4-2} 
\usepackage{graphicx} % needed for figures
\usepackage{float} % locate figures, [H]
\usepackage{dcolumn} % needed for some tables
\usepackage{bm,color}
\usepackage{amssymb}
\usepackage{amsmath}
\usepackage{stmaryrd}
\usepackage{bm}
\usepackage{braket}
\usepackage{extarrows}
\usepackage[colorlinks=true,linkcolor=blue,urlcolor=blue,citecolor=blue]{hyperref}
\usepackage{hyperref}
\usepackage{txfonts}

\usepackage[normalem]{ulem} %makes underlining possible: \uline \uwave \sout
\definecolor{mgreen}{RGB}{1,123,0}

\newcommand{\DQrm}[1]{}

\newcommand{\Yes}{{\color{black}invariant}}
\newcommand{\None}{{\color{black}broken}}

\begin{document}
\title{
Magnetic properties of the quasi two-dimensional centered honeycomb antiferromagnet GdInO$_3$
}

\author{Xunqing Yin}
\affiliation{ Institute of Applied Physics and Materials Engineering, University of Macau, Avenida da Universidade, Taipa, Macau 999078, China}
\author{Yunlong Li}
\author{Guohua Wang}
\author{Jiayuan Hu}
\author{Chenhang Xu}
\author{Qi Lu}
\affiliation{Key Laboratory of Artificial Structures and Quantum Control (Ministry of Education), Shenyang National Laboratory for Materials Science, School of Physics and Astronomy, Shanghai Jiao Tong University, Shanghai 200240, China}
\author{Yunlei Zhong}
\affiliation{ Institute of Applied Physics and Materials Engineering, University of Macau, Avenida da Universidade, Taipa, Macau 999078, China}
\author{Jiawang Zhao}
\author{Xiang Zhao}
\author{Yuanlei Zhang}
\author{Yiming Cao}
\author{Kun Xu}
\author{Zhe Li}
\affiliation{Center for Magnetic Materials and Devices, College of Physics and Electronic Engineering, Qujing Normal University, Qujing 655011, China}
\author{Yoshitomo Kamiya}
\email{yoshi.kamiya@sjtu.edu.cn}
\affiliation{Key Laboratory of Artificial Structures and Quantum Control (Ministry of Education), Shenyang National Laboratory for Materials Science, School of Physics and Astronomy, Shanghai Jiao Tong University, Shanghai 200240, China}
\author{Guo Hong}
\email{ghong@um.edu.mo}
\affiliation{ Institute of Applied Physics and Materials Engineering, University of Macau, Avenida da Universidade, Taipa, Macau 999078, China}
\affiliation{Department of Physics and Chemistry, Faculty of Science and Technology, University of Macau, Avenida da Universidade, Taipa, Macau 999078, China}
\author{Dong Qian}
\email{dqian@sjtu.edu.cn}
\affiliation{Key Laboratory of Artificial Structures and Quantum Control (Ministry of Education), Shenyang National Laboratory for Materials Science, School of Physics and Astronomy, Shanghai Jiao Tong University, Shanghai 200240, China}
\affiliation{Collaborative Innovation Center of Advanced Microstructures, Nanjing 210093, China}
\affiliation{Tsung-Dao Lee Institute, Shanghai Jiao Tong University, Shanghai 200240, China}

\date{\today}
\begin{abstract}

The crystal structure and magnetic property of the single crystalline hexagonal rare-earth indium oxides GdInO$_3$ have been studied by combing experiments and model calculations.
The two inequivalent Gd$^{3+}$ ions form the centered honeycomb lattice, which consists of honeycomb and triangular sublattices.
 The dc magnetic susceptibility and specific heat measurements suggest two antiferromagnetic phase transitions at $T_\textrm{N1}$ = 2.3 K and $T_\textrm{N2}$ = 1.02 K.
An inflection point is observed in the isothermal magnetization curve, which 
can be an indication of
an up-up-down phase with a 1/3 magnetization plateau%
, further supported by our theoretical calculation.
We also observe a large magnetic entropy change originated
from the magnetic frustration in GdInO$_3$.
By considering a classical spin Hamiltonian,
we establish the ground state phase diagram, which suggests that GdInO$_3$ has a weak easy-axis anisotropy and is close to the equilateral triangular-lattice system.
The theoretical ground-state phase diagram may be used as a reference in NMR, ESR, or $\mu$SR experiments in future.
\end{abstract}

\maketitle

\section{Introduction}

Perovskite-type oxides $AB$O$_3$ have been a long-standing research topic in condensed matter physics~\cite{pena2001,li_chemically_2017,kool_properties_2010,varignon_origin_2019}. The intricate competing interplay between the charge, lattice, orbital, and spin degrees of freedom
results rich physical phenomena of potential technological relevance, such as ferromagnetism, antiferromagnetism, ferroelectricity, piezoelectricity, multiferroicity, metal-insulator transition, superconductivity, giant magnetoresistance, and so on~\cite{callaghan_magnetic_1966,cohen_origin_1992,srinivasan_magnetoelectric_2002,cheong_multiferroics_2007,ishiwata_low-magnetic-field_2008,mott_metal-insulator_2004,bednorz_perovskite-type_1988,maeno_superconductivity_1994,moritomo_giant_1996}. Numerous complex structural families have been derived from the distortion of the ideal cubic structure of $AB$O$_3$ perovskites, such as the low-symmetric hexagonal $AB$O$_3$ perovskites exhibiting
 improper geometric ferroelectricity~\cite{cheong_multiferroics_2007,tohei_geometric_2009}. In particular, the hexagonal $R$MnO$_3$ ($R=$ rare earth element) families have been widely studied because of their interesting multiferroic characteristics~\cite{lilienblum_ferroelectricity_2015,valdes_aguilar_origin_2009,song_lattice_2019}, i.e., the coexistence of magnetic ordering and ferroelectricity.

 Mn$^{3+}$ ion can be replaced with the nontransition metal In$^{3+}$ ion. The resulting rare earth indates $R$InO$_3$
 have triggered research interests owing to geometric ferroelectricity~\cite{yu_first_2020,barnita_paul_and_swastika_chatterjee_and_sumana_gop_and_anushree_roy_and_vinita_grover_and_rakesh_shukla_and_a_k_tyagi_evolution_2016,li_laser_2018,paul_geometrically_2017} and the quantum spin liquid (QSL)-like behaviors~\cite{gordon_nonequivalent_2018,kim_spin-liquid-like_2019, kim_spin_2019,clark_two-dimensional_2019}.  GdInO$_3$ has the so-called centered honeycomb lattice structure ($P6_3cm$ space group), which is the same hexagonal structure as the QSL candidate TbInO$_3$ and is characteristic of the $R$InO$_3$ family.
Gd$^{3+}$ has the half-filled $4f$ shell,
as its outmost orbital, with the orbital angular momentum $L = 0$ and the high spin state $S = 7/2$, suggesting a semiclassical and nearly isotropic spin Hamiltonian.
Previous studies have revealed exotic physical properties of GdInO$_3$, such as negative thermal expansion (NTE), spin-lattice coupling and the improper geometric ferroelectricity~\cite{li_laser_2018,paul_geometrically_2017}. However, systematical studies on the ground-state spin configuration have yet been carried out. In the present work, we investigate the
magnetic properties in single crystals of GdInO$_3$ combining dc magnetic susceptibility, heat capacity measurements, and model calculations. We find that the ground state phase diagram of a classical spin model relevant for the materials provides reasonable explanations about our experiments.

\begin{figure}
\centering
\includegraphics[width = 0.48\textwidth] {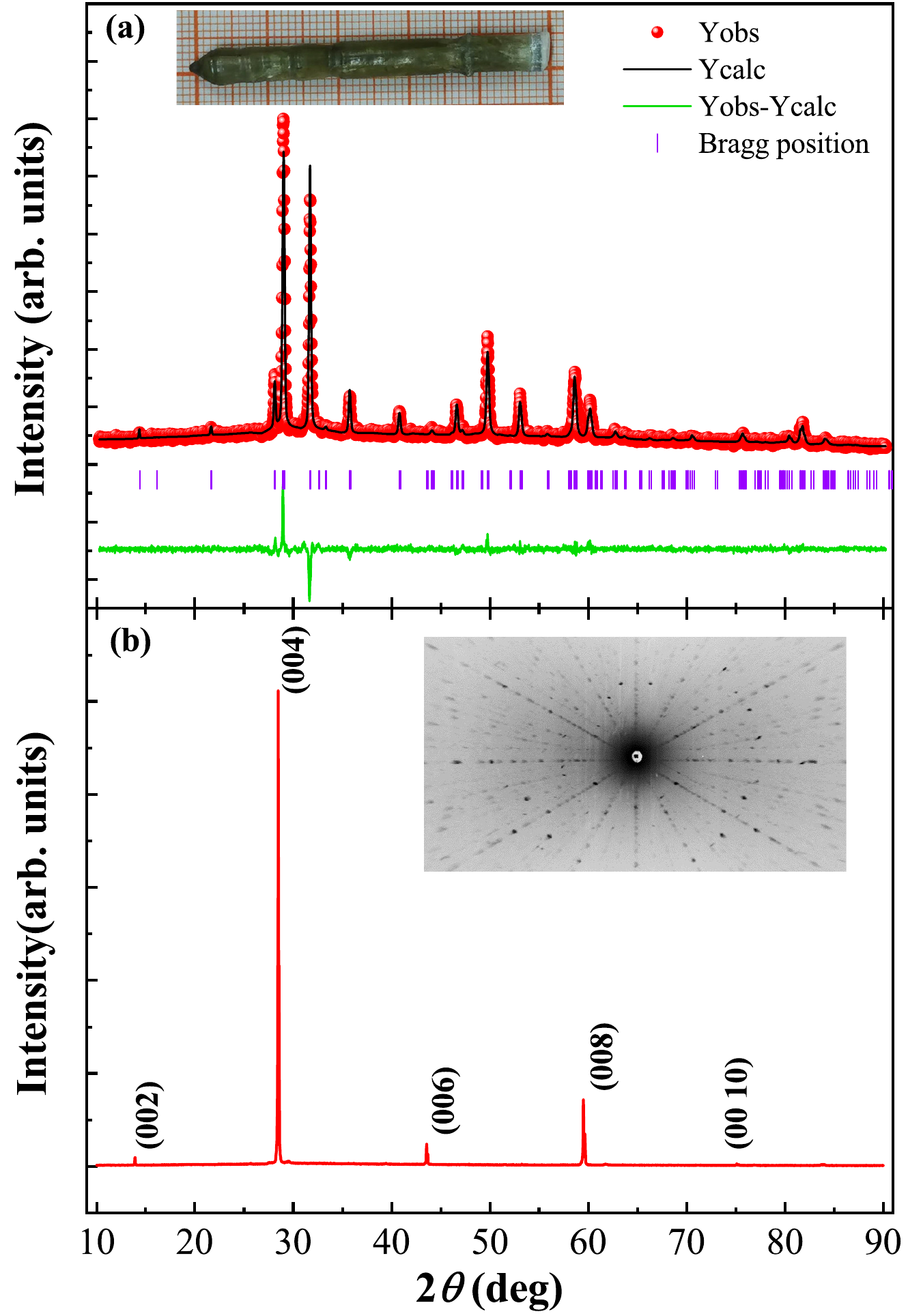}
\caption{(color online) (a) Observed (circles) and calculated (solid lines) XRD patterns  at room temperature. The vertical bars in each panel mark the positions of nuclear Bragg reflections of GdInO$_3$. The lower curves represent the difference between observed and calculated patterns. The inset shows the as-grown boule of GdInO$_3$ single crystal. (b) The XRD pattern of as-grown single crystal along [001] orientation. The inset shows the corresponding Laue photograph.}
\label{XRD}
\end{figure}

\section{Experimental}
\subsection{Method}
 Polycrystalline GdInO$_3$ samples were synthesized by the solid-state reaction method. High-purity In$_2$O$_3$ (99.99\%) and Gd$_2$O$_3$ (99.99\%) were first weighted in the stoichiometric ratio and then mixed. The mixture was pressed into pellet under a pressure of 70 MPa and heated at 950$^{\circ}$, 1150$^{\circ}$and 1350$^{\circ}$ in air, each for 24 hours with intermediate grinding.
 Then the powder was hydrostatically pressed into feed rod (length 60 mm, diameter 8 mm) and seed rod (length 35 mm, diameter 8 mm) and sintered at 1350$^{\circ}$ for 24 hours. A single crystal of GdInO$_3$ was grown using a two-mirror optical floating zone furnace (IRF01-001-05, Quantum Design) with 2$\times$ 650 W halogen lamps. To avoid the severe volatilization of In$_2$O$_3$, the growth was performed under $\sim$ 9 bar oxygen pressure and a flow rate of 0.2 L/min O$_2$. Feed and seed rods were counter-rotated at the same rate of 20 rpm, to improve zone homogeneity. The steady growth speed was 10 mm/hour.

X-ray powder-diffraction was performed with the copper $K_{\alpha}$ = 1.54056 {\AA} radiation at 300 K in X-ray Diffractometer (Rigaku Smartlab 9000W). The single crystal sample was gently ground into powder and then pressed onto a thin flat surface. All powder diffraction data were analyzed by the Fullprof Suites~\cite{rodriguez-carvajal_recent_1993}. Relevant refinement parameters are the wavelength, a scale factor, zero shift, cell constants, shape parameters, asymmetry parameters, preferred orientations, atomic positions and isotropic thermal parameter B, etc. A pseudo-Voigt function was used for the peak profile shape fitting. We also adopt a linear interpolation between automatically-selected background points for the background refinement. The crystallinity and crystallographic orientation were confirmed using a back-reflection Laue X-ray camera.

The magnetization and specific heat measurements were conducted using Physical Property Measurement System (PPMS Dynacool, Quantum Design).

\begin{table}
\centering
\caption{Refined structural parameters with $P6_3cm$ space group symmetry for GdInO$_3$ at room temperature.}
\label{table0}
\begin{tabular} {ccccccccc}
\hline
\multicolumn{3}{c}{Cell Parameters}\\
 $a$({\AA})  & $c$({\AA})  &  $V$({\AA}$^3$)\\
 6.3433(3) & 12.3320(1) & 429.760(3) \\
 \hline
\multicolumn{4}{c}{Reliability factors} \\
$R_\texttt{p}$ &  $R_\texttt{wp}$ & $R_\texttt{exp}$ & $\chi^2$ \\
5.04  & 6.67 & 3.02 & 4.87 \\
\hline
Atom & Wyckoff  &  \emph{x} & \emph{y} & \emph{z} & $B$ ({\AA}$^2$) \\
~&positions\\
Gd1 & 2a & 0.00000 &  0.00000 & 0.2741(8) & 4.64(8)  \\
Gd2 & 4b & 0.33333 &  0.66667 & 0.2459(7) & 4.64(8)  \\
In1 &6c & 0.3350(3) & 0.00000 & 0.00000  & 2.44(3)  \\
O1  &6c& 0.3121(7) & 0.00000 & 0.1552(9) & 5.00(0) \\
O2  &6c & 0.6188(3) & 0.00000 & 0.3153(6) & 5.00(0) \\
O3  &2a& 0.00000 & 0.00000 & 0.4801(0) & 5.00(0) \\
O4  &4b& 0.33333 & 0.66667 &  0.0260(9) & 5.00(0) \\
\hline
\end{tabular}
\end{table}

\subsection{Structure characterization}

\begin{figure*}
\centering
\includegraphics[width = 0.96\textwidth] {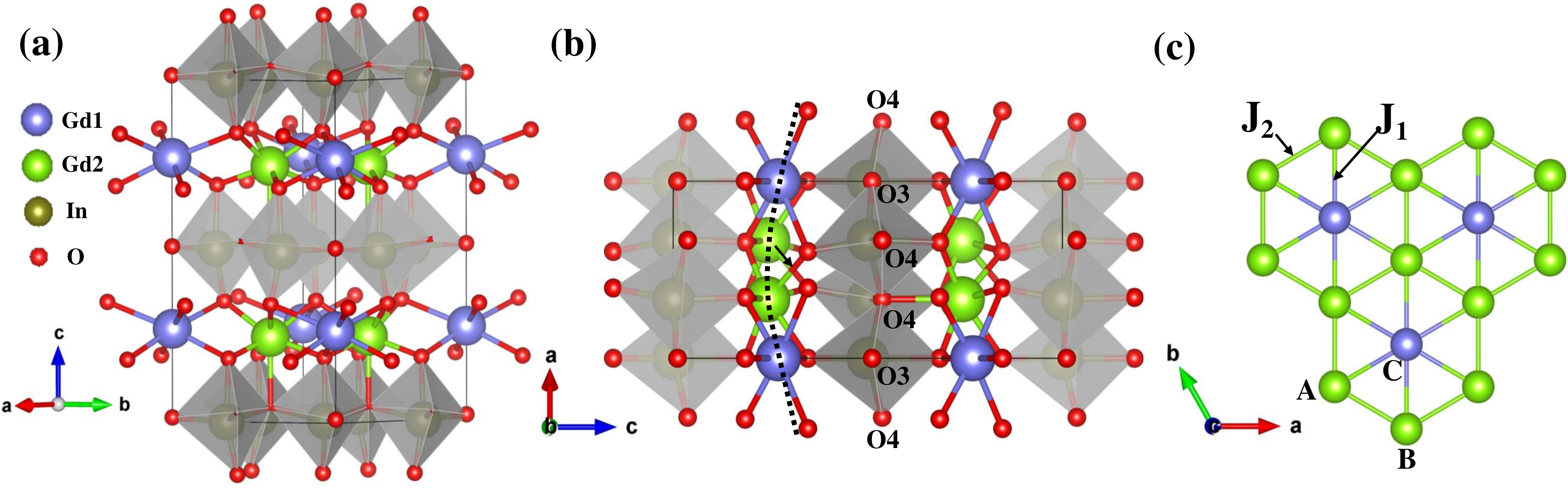}
\caption{(color online) (a) Crystal structure ($P6_3cm$) with one unit cell of GdInO$_3$. (b) Crystal structure with one unit cell of GdInO$_3$ along the [010] projection, the curved dotted line represents the arc-like alignment of two inequivalent atomic sites of Gd ions. The three planar oxygen ions(O3, O4, O4) are also labeled.
(c) Two inequivalent atomic sites of Gd ions form the stuffed honeycomb lattice along the [001] projection. $J_1$ is the nearest neighbor (NN) spin coupling constant between Gd1 and Gd2, $J_2$ is the NN spin coupling constant between Gd2. The sublattice labels A and B represent Gd2 and C represents Gd1.
}
\label{Cell}
\end{figure*}

Figure \ref{XRD}(a) shows the X-ray diffraction (XRD) pattern of the single crystal powder of GdInO$_3$ performed at 300 K. Rietveld refinement was performed under the hexagonal space group $P6_3cm$ using Fullprof Suites. The key structural features of GdInO$_3$ are in reasonable agreement with the previous measurements~\cite{li_laser_2018,paul_geometrically_2017}.
The agreement between the experiments and the simulated profile is excellent. All the observed Bragg peaks can be well indexed and no extra peaks were detected.
The refined results
are listed in Table \ref{table0}.
Figure \ref{XRD}(b) shows the XRD patterns of GdInO$_3$ single crystal,
where five
reflections $(0 0 l)$ ($l = 2,4,6,8,10$) are observed.
The sharp diffraction peaks
and the nice Laue backscattering diffraction pattern in the ab-plane [the inset of Fig.~\ref{XRD}(b)],
indicates high quality of our single crystal.

The crystal structure in a unit cell is shown in Fig.~\ref{Cell}(a). Gd1 and Gd2 are two inequivalent Gd atoms with Wyckoff positions 2a and 4b, respectively. The hexagonal structure consists of tilted InO5 bipyramids with two apical (O1, O2) and three planar oxygen ions (O3, O4, O4)
(shown in Fig. \ref{Cell}(b)). 
The corner-linked layered InO5 bipyramids are separated by alternating Gd layers. The two
inequivalent atomic positions of Gd ions form the arc-like arrangement viewed along the [010] projection
[Fig.~\ref{Cell}(b)].
When viewed along the [001] projection, the two inequivalent atomic positions of Gd ions form the
centered honeycomb lattice, which consists of a honeycomb lattice with a superimposed triangular lattice located at the center of each hexagon [Fig.~\ref{Cell}(c)].
The displacements of the Gd1 and Gd2 sites and the associated tilting and distortion of InO$_5$ bipyramids create electric
dipoles, which are responsible for occurrence of
the improper geometric ferroelectricity~\cite{li_laser_2018}
in GdInO$_3$.

\subsection{dc magnetic susceptibility}

\begin{figure}
\centering
\includegraphics[width = 0.48\textwidth] {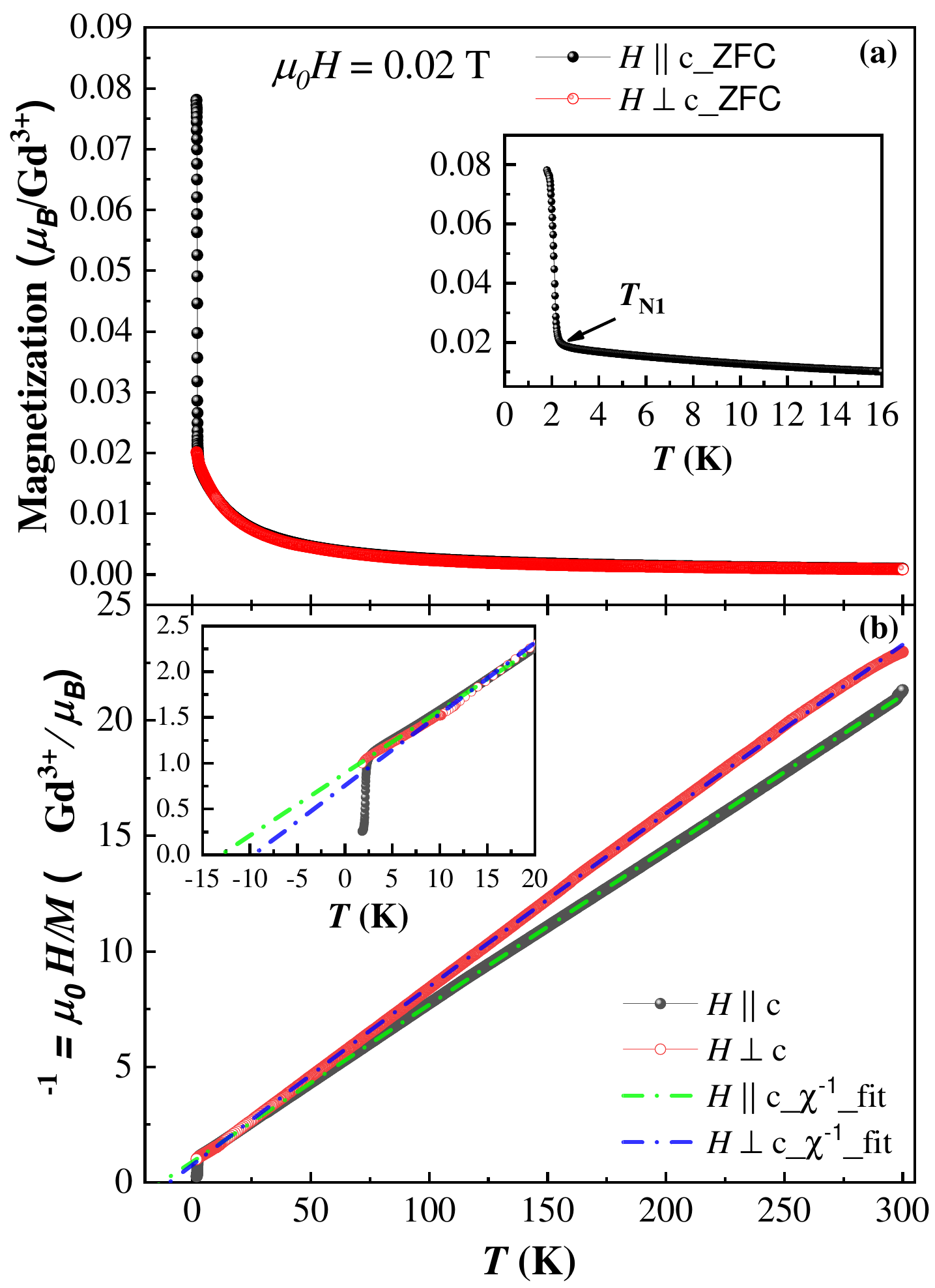}
\caption{(color online)  (a) Temperature-dependent magnetization for $\mathbf{H} \parallel c$ and H $\perp$ c of  GdInO$_3$ with ZFC process under 200 Oe. The inset shows an expansion of the magnetization over the temperature range 0 K to 16 K where magnetic transition $T_\textrm{N1}$ appears. (b) Inverse magnetic susceptibility $\chi^{-1}$ deduced from the ZFC magnetization. The dashed and dash-dotted lines are fit to the data with a CW law as described in the text.}
\label{MT}
\end{figure}

Figure~\ref{MT}(a) shows the temperature dependence of
ZFC and FC dc magnetic susceptibility $M(T)$
for parallel and perpendicular to the crystallographic c axis of GdInO$_3$
measured at $\mu_0 H = 0.02$ T.
The $M(T)$ for $H\perp c$ shows paramagnetic behavior
down to 1.8 K.
On the other hand, $M(T)$ for $H \parallel c$
shows an upturn of the $M(T)$
curve at $\sim$ 2.3 K, as
shown in the inset of Fig. \ref{MT}(a).
This magnetic transition has not been reported previously and we label the transition temperature as $T_\textrm{N1}$.

The inverse magnetic susceptibility
${\chi}^{-1} = {\mu}_{0}H/M$
is shown in Fig. \ref{MT}(b). A Curie-Weiss (CW)
analysis of the paramagnetic state of GdInO$_3$ is
performed by using
\begin{equation}\label{1}
  \chi(T)
  = \chi_0 + \frac{N_{A}M_\text{eff}^{2}}{3 k_\text{B}(T - \theta_\text{CW})},
\end{equation}
where $\chi_0$ is a temperature-independent term, $\theta_\textrm{CW}$ is the CW temperature,
$M_\textrm{eff}$ is the effective paramagnetic moment,
$N_\textrm{A}$ is the Avogadro constant, and
$k_\textrm{B}$ is the Boltzmann constant.
The derived effective moments for $\mathbf{H} \parallel c$ and $\mathbf{H} \perp c$ are $M_\text{eff}$ = 7.55 $\mu_{B}$ and 8.08 $\mu_{B}$ per Gd$^{3+}$, respectively, which are close to the theoretical prediction 7.94 $\mu_{B}$ for a free Gd$^{3+}$ ion. The
CW temperature for $\mathbf{H} \parallel c$ and $\mathbf{H} \perp c$ are $\theta_\text{CW}=$ -13.06 K and -9.66 K, respectively.
The negative CW temperature suggests that the transition at $T = T_\textrm{N1}$ is an onset of an antiferromagnetic ordering.
The frustration parameter is
$\left| \frac{\theta_\textrm{CW}}{T_\textrm{N1}} \right| \sim 6.0$,
which is rather large, indicating strong spin frustration in GdInO$_3$.

\begin{figure}
\centering
\includegraphics[width = 0.48\textwidth] {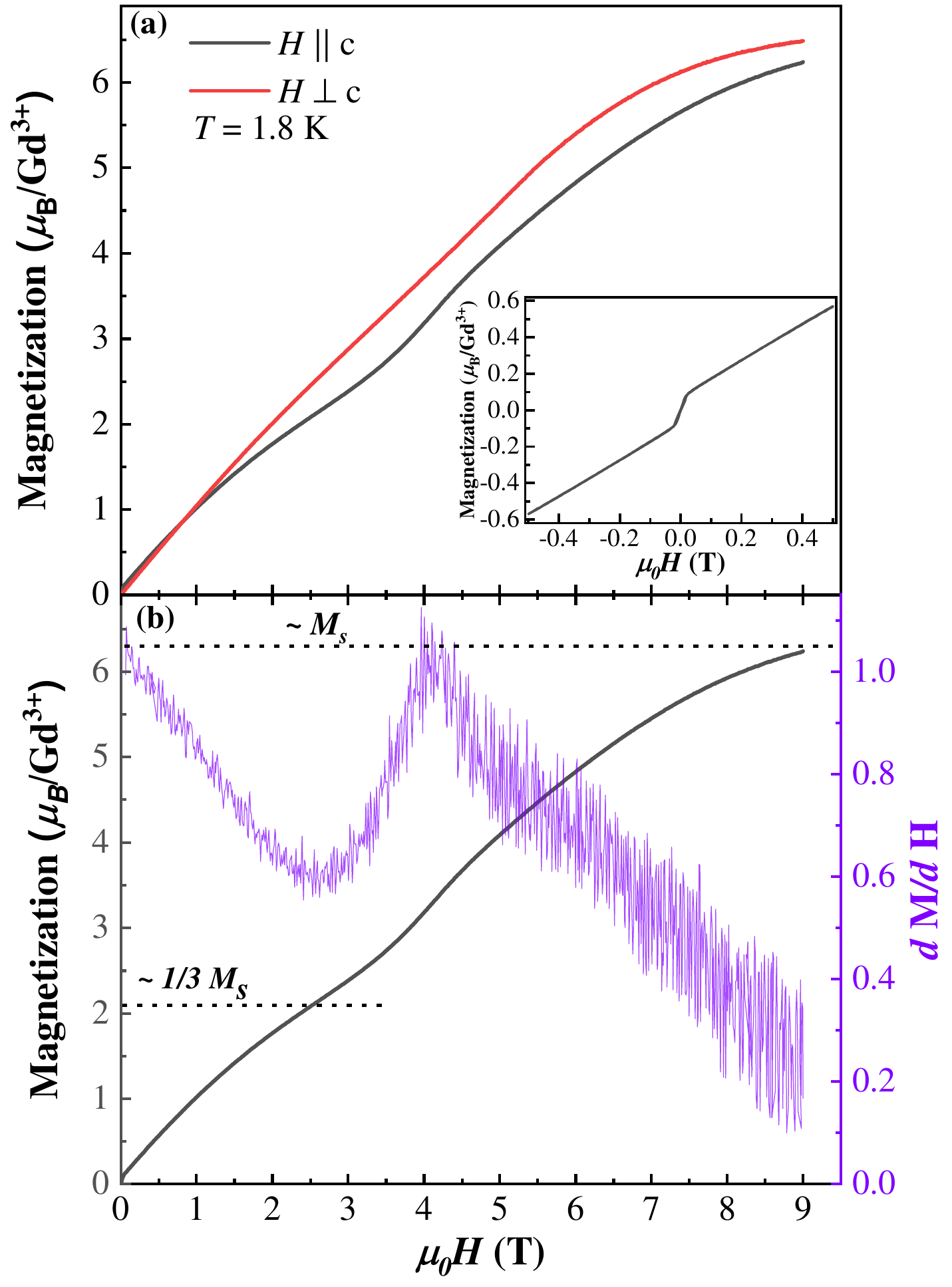}
\caption{(color online) (a) Isothermal magnetization curves for $\mathbf{H} \parallel c$ and H $\perp$ c obtained at 1.8 K.
A very tiny loop of magnetization curve for $\mathbf{H} \parallel c$ is shown in the inset.
(b) Isothermal magnetization curves for $\mathbf{H} \parallel c$ obtained at 1.8 K(black curves) and its first derivative d$M$/d$H$(purple curves). The upper dotted line represents the saturation magnetic moment, the lower dotted line indicates the transitions close to 1/3 of $M_s$.
}
\label{MHloop}
\end{figure}

The isothermal magnetization curves $M(H)$ for $\mathbf{H} \parallel c$ and $\mathbf{H} \perp c$ at 1.8 K are shown in Fig. \ref{MHloop}(a). The different behaviors for
$\mathbf{H} \parallel c$ and $\mathbf{H} \perp c$ indicate
magnetic anisotropy. We observe a very tiny hysteresis loop for $\mathbf{H} \parallel c$, suggesting the existence of small magnetic moment
along the $c$ axis.
The individual $M(H)$ curve and its first derivative (d$M$/d$H$) at 1.8 K along $c$ axis is shown in Fig. \ref{MHloop}(b).
One notable inflection point emerged around 2.6 T, the corresponding magnetization 
value is about 1/3 of the saturation magnetic moment ($M_s$).
The inflection point can be more clearly seen in d$M$/d$H$ curve as a valley-peak characteristics.
This anomaly may indicate either a tiny 1/3 magnetization plateau, as observed in other triangular lattice antiferromagnets (TLAFs)~\cite{lee_magnetic_2014,shirata_experimental_2012,
lee_series_2014,ahmed_multiple_2015,smirnov_triangular_2007}, or simply a magnetization kink.
Our model calculations support the 1/3 magnetization plateau scenario under a set of reasonable assumptions, which will be discussed in the theory section (see Sec.~\ref{sec:theory}).

\begin{figure}
\centering
\includegraphics[width = 0.48\textwidth] {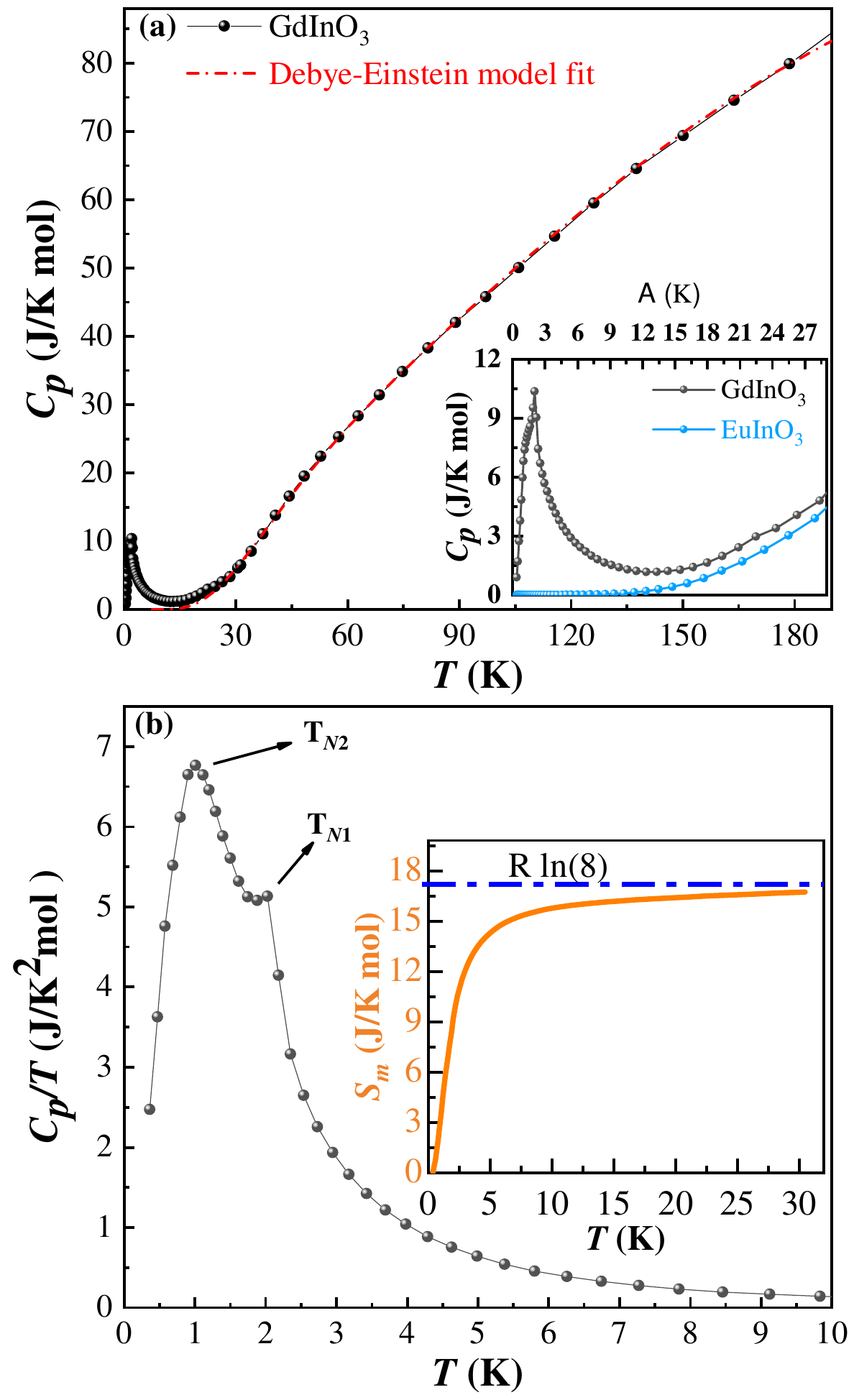}
\caption{(color online)
(a) Specific heat $C_p$ (solid circles) as a function of temperature in zero field and the Debye and Einstein terms for the lattice heat capacity fit (dash dot lines), the inset show the $C_p(T)$ curves of GdInO$_3$ and EuInO$_3$ in the low temperature range from 0.35 K to 30 K.
(b) $C_p(T)/T$ curves of GdInO$_3$ in the low temperature range from 0.35 K to 30 K. The two black arrows indicate the two successive magnetic phase transitions $T_\textrm{N1}$ and $T_\textrm{N2}$. The inset shows the full magnetic ions Gd$^{3+}$ contribution to the entropy, as indicated by the blue dotted line positioned at R ln(8).
}
\label{Cp}
\end{figure}

\subsection{Heat Capacity}

The temperature dependent specific heat $C_p(T)$ in zero field is shown in Fig. \ref{Cp}. $C_p(T)$ at high temperature is dominated by phonons and the magnetic contribution becomes prominent below $\sim 15$ K.
Two successive sharp peaks are observed in $C_p(T)/T$ at $T \sim$ 1.05 K and $\sim$ 2.2 K [Fig. \ref{Cp}(b)].
Comparing with magnetic measurements, the first transition at $\sim$ 2.2 K is the antiferromagnetic transition. We think the second transition at lower temperature could be another magnetic transition.
The lattice specific heat contribution has been estimated by fitting the high-temperature data above 30 K.
We use the Debye-Einstein model
$
C_\textrm{lattice}(T) = C_\textrm{Debye} + C_\textrm{Einstein}
$
for the lattice heat capacity~\cite{tari_specific_2003}:
\begin{align}
\label{5}
C_\textrm{Debye} &
= 9
n_D
R\left(\frac{T}{\Theta_D}\right)^3\int_{0}^{\Theta_D / T}\frac{x^4 e^x}{(e^x - 1)^2}\,dx,
\\
C_\textrm{Einstein} &=
%\sum
\sum_i 3
n_{E_{i}}
%n
R\left(\frac{\Theta_{E_{i}}}{T}\right)^2\frac{ e^{\Theta_{E_{i}}/T}}{(e^{\Theta_{E_{i}}/T} - 1)^2},
\end{align}
where $R$ denotes the gas constant,
$\Theta_{D}$ and $\Theta_{E_{i}}$ are the Debye and the $i$-th Einstein temperatures, respectively,
and $n_D$ and $n_{E_{i}}$ are integer weight factors.
Here, $n_D$ and $n_{E_{i}}$ must satisfy $n_D + \sum_i n_{E_{i}} = n$ where $n = 5$ is the number of atoms in a forum unit.
The best fit for the high temperature specific heat under zero field
can be obtained by assuming a single Debye branch ($n_D = 1$, $\Theta_D$ = 740 K) and three Einstein branches,  $(n_{E_1},n_{E_2},n_{E_3}) = (1,1,2)$ with $(\Theta_{E_1},\Theta_{E_2},\Theta_{E_3}) = (143\,\mathrm{K},214\,\mathrm{K},583\,\mathrm{K})$.

Since GdInO$_3$ is a insulator, the electronic contribution to the total heat capacity can be neglected at low temperature. By using the
low temperature lattice heat capacity of nonmagnetic insulator EuInO$ _3$ [inset of Fig. \ref{Cp}(a)]
assuming it has similar lattice heat capacity as GdInO$_3$,
we extract the magnetic heat capacity $C_m$ of GdInO$_3$. The entropy change $S_m$ is then calculated by $\Delta S_M(T) = \int_{0}^{T}\frac{C_m(T)}{T}\,dT $ [inset of Fig. \ref{Cp}(b)]. The entropy change is about 16.73 J K$^\texttt{-1}$mol$^\texttt{-1}$, which is close to the expected value,
$R$ $\ln 8 = 17.29$ J K$^\textrm{-1}$mol$^{-1}$, for a spin-7/2 system.

\begin{figure}
\centering
\includegraphics[width = 0.48\textwidth] {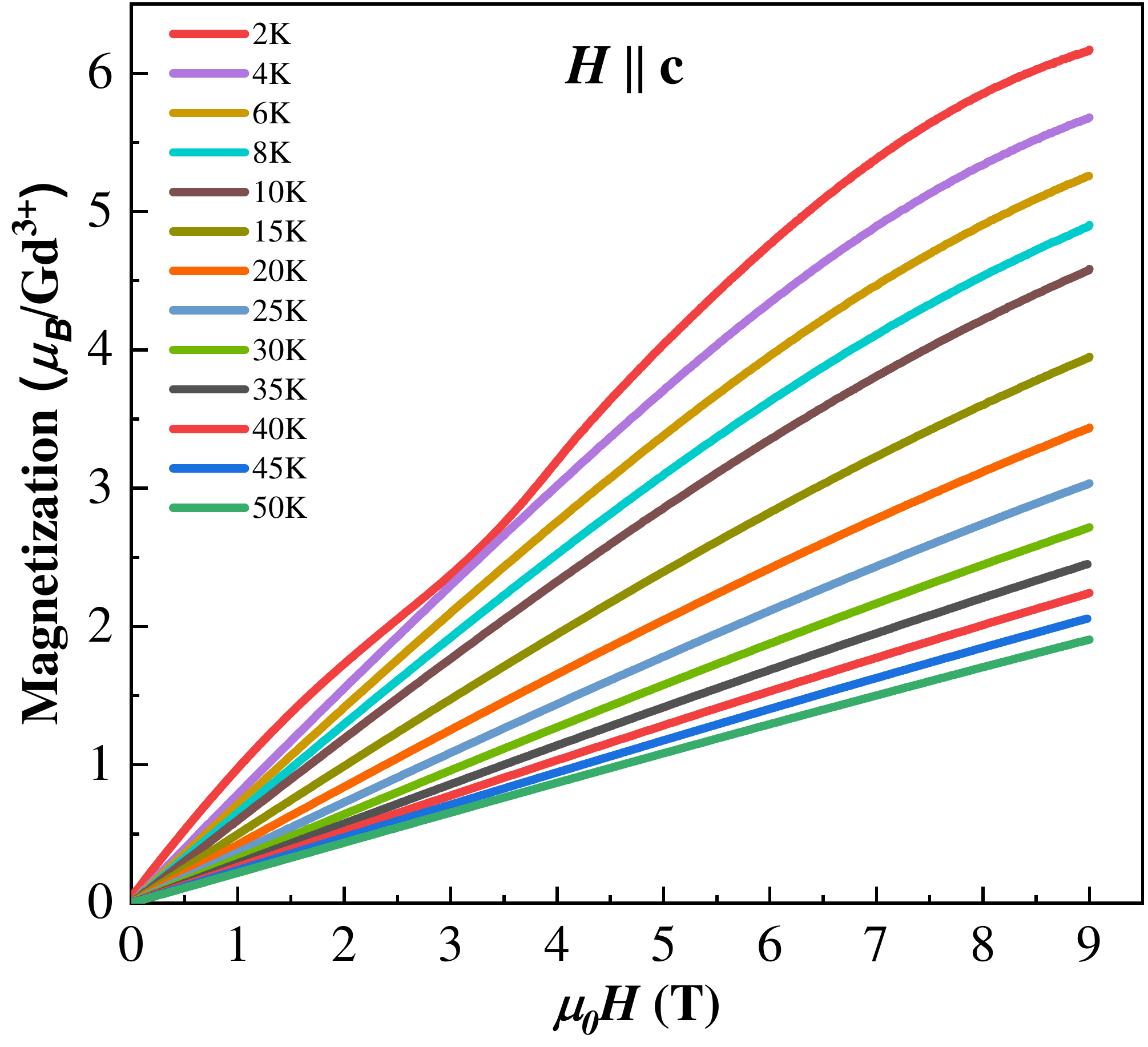}
\caption{(color online)  Isothermal magnetization curves obtained in the temperature range from 2 K to 50 K and under magnetic fields 0 to 9 T.}
\label{MH}
\end{figure}

\begin{figure}
\centering
\includegraphics[width = 0.48\textwidth] {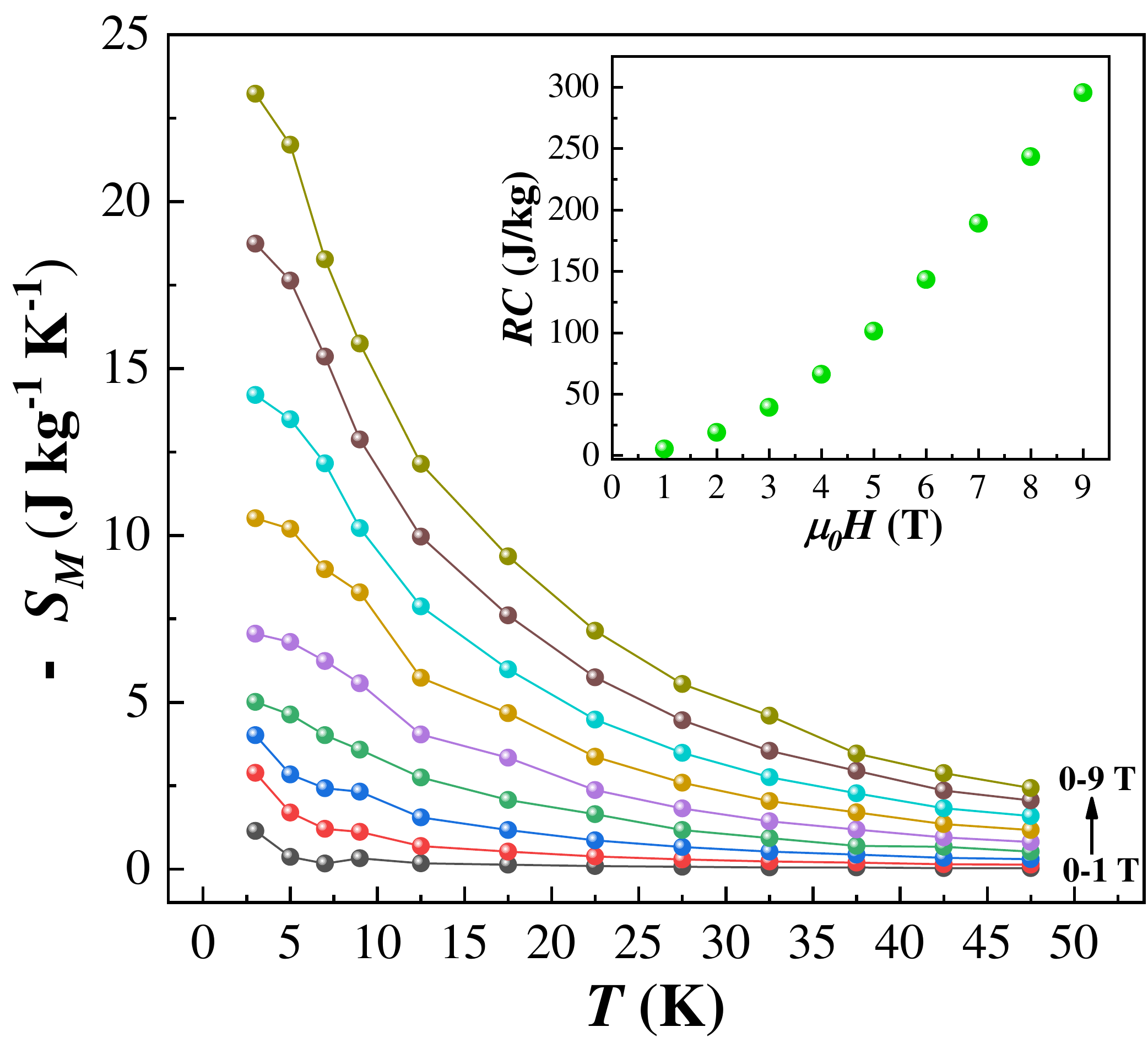}
\caption{(color online)
Temperature variation of magnetic entropy change
-$\Delta S_{\textrm{M}}$ for GdInO$_3$ calculated from the magnetization data. The inset shows RC as a function of magnetic field.}
\label{RCP}
\end{figure}

\subsection{Magnetocaloric effect}

The field dependence of the magnetization from 2 K to 50 K are shown in Fig. \ref{MH}. The magnetocaloric effect (MCE) describes the temperature change of a magnetic material under adiabatic conditions through the application or removal of an external magnetic field. This effect is particularly pronounced at temperatures and magnetic fields corresponding to magnetic phase transitions. The MCE measurement is a powerful and widely used tool for investigating  magnetic refrigeration
materials. Compared with conventional systems based on the vapour–gas cycle techniques, alternative techniques utilizing MCE can be efficient in low-energy consumption and safe to the environment~\cite{liu_giant_2012,shen_recent_2009}.
Recently, frustrated magnetic materials begin to be recognized for its great potential in enhancing the cooling power
related to the presence of a macroscopic number of soft modes below the saturation field~\cite{pakhira_magnetic_2017,sahoo_influence_2018,
zhitomirsky_enhanced_2003}.
We evaluate the magnetic entropy change of GdInO$_3$ for $\mathbf{H} \parallel c$ using the Maxwell relation:
\begin{equation}\label{2}
  \Delta S_\text{M}(T,H)
  = \int_{0}^{H}\mu_0\left(\frac{dM}{dT}\right)_{H}\, dH,
\end{equation}
The temperature dependence of -$\Delta S_{\textrm{M}}$ under magnetic fields up to 9 T is shown in Fig.~\ref{RCP}. -$\Delta S_{\textrm{M}}$ increases with the increase of applied magnetic field. With maximum field of 9 T,  -$\Delta S_{\textrm{M}}$ increases $\sim$  23.19 J kg$^{-1}$K$^{-1}$  at 2 K.
The refrigeration capacity (RC) of a magnetocaloric material can be calculated by

\begin{equation}\label{4}
  RC = -\int_{T_1}^{T_2}\Delta S_\text{M}(T)\,dT,
\end{equation}
where $T_1$ - $T_2$ = $\delta T_\textrm{FWHM}$ and the $\delta T_\textrm{FWHM}$ is its full width at half maximum of -$\Delta S_{\textrm{M}}$. 
The RC as a function of different field changes is depicted in the inset Fig.~\ref{RCP}. 
Its RC reaches $\sim$  200 J kg$^{-1}$ (at $\Delta \mu_0H$ = 7 T) and $\sim$  295 J kg$^{-1}$ (at $\Delta \mu_0H$ = 9 T). 
GdInO$_3$'s RC is comparable to that in other Gd-based magnetic materials with and without significant frustrated interactions. For example, RC = 325 J kg$^{-1}$ for $\Delta \mu_0H$ of 7 T in  SrGd$_2$O$_4$~\cite{jiang_magnetization_2018}, RC = 125 J kg$^{-1}$  for $\Delta \mu_0H$ of 7 T in RuSr$_2$GdCu$_2$O$_8$~\cite{midya_giant_2014}, RC = 400 J kg$^{-1}$ for $\Delta \mu_0H$ of 8 T in GdMnO$_3$~\cite{wagh_low_2015}. Considering its relatively large RC among these other Gd-based materials, the inappreciable field/thermal hysteresis and the weak anisotropy, GdInO$_3$'s  could be a candidate for a cryogenic magnetic refrigeration material.

The relatively large magnetic entropy change above 5 T near liquid temperature
in GdInO$_3$ may result from the field-induced paramagnetic state.
A large magnetic entropy change has been also observed in
SrGd$_2$O$_4$~\cite{jiang_magnetization_2018}, but it is distinct from unfrustrated systems where the magnetic entropy change tends to grow rapidly at low fields and saturate at high fields.
This distinct behavior can possibly be attributed to
the presence of a macroscopic number of soft modes
due to frustration below the saturation field~\cite{zhitomirsky_enhanced_2003}.

\section{%
Theory
\label{sec:theory}
}
In the following, we 
discuss our theoretical standpoint about GdInO$_3$ using a classical spin Hamiltonian relevant for this material. As mentioned before, Gd$^{3+}$ ions in GdInO$_3$ (space group: $P6_3cm$) comprises three sublattices $A$--$C$, with sublattices $A$ and $B$ (Gd2) forming the honeycomb network and sublattice $C$ (Gd1) in the center of every hexagon, which corresponds to a $\sqrt{3}\times\sqrt{3}$ deformation of the  triangular lattice. 
A free Gd$^{3+}$ ($4f^7$) carries $S = 7/2$ with a quenched orbital moment $L = 0$, for which a nearly isotropic $g$-tensor is expected, as in fact seen in our experiments (Fig.~\ref{MT}). As usual, a second-order perturbative effect due to the spin-orbit coupling can give rise to single-ion anisotropy. Indeed, the measured isothermal magnetization curves for $H \parallel c$ and $H \perp c$ behave differently, especially in high fields (Fig.~\ref{MHloop}).
Thus, we consider the following classical spin Hamiltonian, which is expected to be adequate for $S = 7/2$ spins:
\begin{align}
    H = \sum_{\langle{ij}\rangle}J_{ij} \mathbf{S}_i \cdot \mathbf{S}_j - D\sum_i \left(S_i^z\right)^2
    - \mathbf{h} \cdot \sum_i \mathbf{S}_i,
    \label{eq:H}
\end{align}
where $\mathbf{S}_i$ is a vector spin with unit length at site $i$. We define $J_{ij}$, $D$, and $\mathbf{h}$ as incorporating the factors of $S^2$, $S^2$, and $g\mu_B S$, respectively, and assume the isotropic $g$ tensor. The model is $U(1)$ ($Z_2$) invariant for $\mathbf{h} \parallel c$ ($\mathbf{h} \perp c$) and O(3) invariant for $D = h = 0$. We consider two interactions that are distinguished by symmetry: $J_1$ between sublattices $AC$ or $BC$ and $J_2$ between sublattices $AB$ [Fig.~\ref{Cell}(c)].
We restrict our attention mostly to the frustrated case with antiferromagnetic coupling $J_1,~J_2 > 0$.

Here we comment in passing on the past studies about the centered honeycomb lattice Hamiltonian~\eqref{eq:H}. In the literature, this model has been studied in relation with $ABX_3$ ($A$ is an alkali metal, $B$ is a transition metal, and $X$ is a halogen atom) materials with the $\sqrt{3}\times\sqrt{3}$ deformation, e.g., RbFeBr$_3$, RbMnBr$_3$, and KNiCl$_3$ (see Ref.~\onlinecite{Collins1997} and references therein). However, these materials are quasi-one-dimensional easy-plane antiferromagnets. Because of the anisotropy and the strong intrachain antiferromagnetic interaction [i.e., interlayer interactions neglected in Eq.~\eqref{eq:H}], with the latter several orders of magnitude larger than $J_1$ and $J_2$, the ordered moments in these materials are in the $ab$ plane in the zero field and the off-plane canting costs huge energy. For this reason, the main focus in the past studies was in the phase diagram for $D \le 0$ and $\mathbf{h} \perp c$ or in the perturbative response to a very small field for $\mathbf{h} \parallel c$~\cite{Tanaka1989,Kawamura1990,Plumer1991,Zhang1993,Zhitomirsky1995}. In contrast, GdInO$_3$ is quasi-two-dimensional and nearly isotropic. Hence, to provide a comprehensive viewpoint, we study the entire magnetization process for both $D>0$ and $D<0$ including $\mathbf{h} \parallel c$ and $\mathbf{h} \perp c$.

A previous density functional calculation suggested that interlayer interactions are dominantly \textit{ferromagnetic} and small~\cite{Gordon2018}. On this basis, we consider a two-dimensional lattice by neglecting them while the phase diagram in a quasi-two-dimensional lattice may be discussed based on our results.
The density functional study also predicted $J_1/J_2 \approx 1/3$~\cite{Gordon2018}.
Generally, the modulation $J_1 \ne J_2$ tends to reduce the frustration, as can be inferred from the two limiting cases, $J_1 = 0$ or $J_2 = 0$, which are unfrustrated. Nevertheless, as it turns out, many new phases that are absent in the well-known case $J_1 = J_2$ appear as intermediate phases.
With details about the theoretical magnetic phase diagram put in Appendix~\ref{app:theory}, we discuss below the main conclusions relevant for the material.

The most important experimental clue available at present is the small anomaly in $M(H)$ near 1/3 of the saturation moment for $\mathbf{H} \parallel c$ and the absence thereof for $\mathbf{H} \perp c$ (Fig.~\ref{MHloop}). In addition,
$M(H)$ for $\mathbf{H} \parallel c$ exhibits a small hysteresis loop while one for $\mathbf{H} \perp c$ does not, which indicates a small net magnetic moment along the $c$ axis in the zero field limit.
The subtle anomaly for $\mathbf{H} \parallel c$, though not conclusively solely based on experimental observations, could be indicative of a narrow magnetization plateau, since it is reminiscent of $M(H)$ in the classical triangular-lattice Heisenberg model, where the up-up-down state is selected by the thermal order-by-disorder mechanism and further stabilized by easy-axis anisotropy~\cite{Kawamura1985,Miyashita1986}.
According to our analysis and previously known results, the 1/3 magnetization plateau can be realized at the classical level at $T = 0$ for
(i) $J_1 < J_2$ and $D > 0$,
(ii) $J_1 = J_2$ and $D > 0$,
(iii) $J_1 > J_2$ and $D \ge 0$,
(iv) $J_1 > J_2$ and $D^{\ast\ast} \le D < 0$,
and
(v) $J_1 > J_2$ and $\max(D^{\ast},D^{\ast\ast\ast}) < D < D^{\ast\ast}$,
where 
$D^{\ast\ast} = [\frac{3}{4}(J_1/J_2)^2 - \frac{9}{4}(J_1/J_2) + \frac{3}{2}]J_2$
and
$D^{\ast\ast\ast} = -(9/4)J_1 + 3J_2$ 
[see Fig.~\ref{fig:Mz_large-J:UUD-C}(a) in Appendix \ref{app:subsec:magnetization:c}]. 
In addition, small quantum effects as well as thermal entropic effects are expected to enhance the plateau, though probably not very significantly~\cite{Chubukov1991,Kawamura1985,Koutroulakis2015,Kamiya2018}.
Among the above cases (i)--(v), only the case (iii) is consistent with the tiny net magnetic moment along the $c$ axis in the zero-field limit of the centered honeycomb lattice ($J_1 \ne J_2$).
Roughly speaking, the feature due to the plateau becomes obscure for $J_1 \approx J_2$ [Figs.~\ref{fig:Mz_small-J} and \ref{fig:Mz_large-J} in Appendix~\ref{app:theory}]. Although many other factors (such as disorder and thermal effects) could also contribute to the smearing of an anomaly, we could probably argue that GdInO$_3$ is much closer to the equilateral triangular-lattice system than previously proposed by a density functional calculation~\cite{Gordon2018}.
We also note that the comparison about the magnetization process for $\mathbf{H} \perp c$ is consistent with the scenario of $J_1 \gtrsim J_2$ with $D > 0$.
In the zero-field ground state phase diagram shown in Fig.~\ref{fig:GS} (Appendix~\ref{app:theory}), Phase 6 is the only possibility of a noncollinear order with zero net magnetic moment in the $ab$ plane.
Furthermore, the transverse magnetization curve for $J_1 \gtrsim J_2$ and $D > 0$ exhibits a much narrower 1/3 magnetization plateau than the longitudinal one for the same parameter set; see the demonstration for $(D/J_2, J_1/J_2) = (0.05,1.1)$ in Fig.~\ref{fig:Mx}(c) (Appendix~\ref{app:theory}). We expect that the more obscure anomaly for $\mathbf{H} \perp c$ would be even more easily smeared out by thermal or disorder effects.

\section{Discussions and summary}

In summary, we have reported the magnetization and specific-heat measurements on
the single-crystal GdInO$_3$, where semiclassical $S=7/2$ magnetic moments due to $Gd^{3+}$ ions form the centered honeycomb lattice. Our specific-heat measurements have revealed two phase transitions
at 1.02 K and 2.3 K, in zero field.
At 1.8 K, an inflection point is observed near 1/3 $M_s$ in $M(H)$ curve for $\mathbf{H} \parallel c$.
Also, a tiny hysteresis loop is observed for $\mathbf{H} \parallel c$ at low temperature.
In GdInO$_3$, the coupling between two nearest neighbor Gd$^{3+}$ ions is antiferromagnetic. With the easy-axis anisotropy, two antiferromagnetic transitions were theoretically predicted and indeed observed in some triangular-lattice materials, such as Rb$_4$Mn(MoO$_4$)$_3$~\cite{Yamamoto2015,capriotti_long-range_1999,seabra_phase_2011}. Here, the $C_3$ symmetry of the triangular lattice is first broken to form the up-up-down phase, which then undergoes an additional transition into the so-called Y phase at a lower temperature ~\cite{lee_magnetic_2014,shirata_experimental_2012,lee_series_2014}. We think that two antiferromagnetic transitions in GdInO$_3$ are likely to have a similar origin.

Our model calculations suggest that a Hamiltonian with weak easy-axis anisotropy $D > 0$ and $J_1 \gtrsim J_2$ is consistent with the experiment.
Compared with TbInO$_3$~\cite{gordon_nonequivalent_2018,kim_spin-liquid-like_2019, kim_spin_2019,clark_two-dimensional_2019},
a candidate of a QSL with the same crystal structure, GdInO$_3$ seems to behave much more classically, though with the possibility of multiple phase transitions induced by both magnetic field and temperature.
To gain more detailed information on the realized spin configuration,
additional experimental studies, such as NMR, ESR, or $\mu$SR, would be required and our analysis of the ground-state spin structures can be used as a reference.

\appendix

\section{%
Details of the theoretical analysis
\label{app:theory}
}
We discuss the classical ground state phase diagram of the Hamiltonian~\eqref{eq:H}.
After discussing the ground state phase diagram in zero field (Appendix~\ref{app:subsec:zero-field}), we discuss magnetization processes for $\mathbf{H} \parallel c$ (Appendix~\ref{app:subsec:magnetization:c}) and $\mathbf{H} \perp c$ (Appendix~\ref{app:subsec:magnetization:ab}).

\subsection{%
Ground state phase diagram in zero field
\label{app:subsec:zero-field}
}

\begin{figure}
    \begin{center}
    \includegraphics[width=\columnwidth,bb=0 0 564 544]{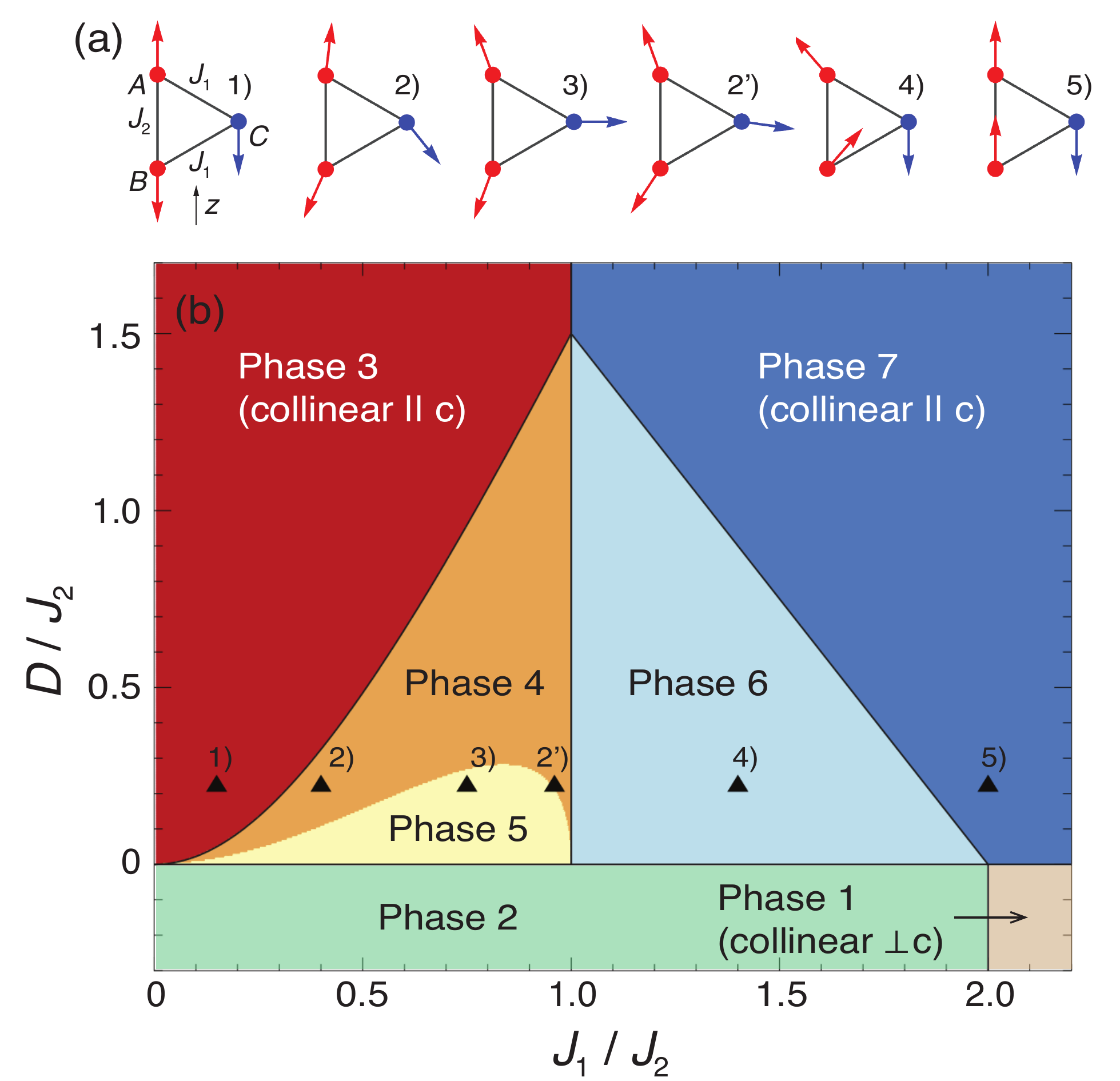}
\end{center}
\caption{
  \label{fig:GS}
    (a) Ground state spin configurations at the selected points (triangles) in the $D>0$ region and (b) the ground state phase diagram in zero field. The states shown in (a) represent the parameter points indicated by triangles in panel (b). Phase 1 is (Phases 3 and 7 are) collinear in the $ab$ plane (along $c$). Other phases are noncollinear and coplanar; Phase 2 is (Phases 4--6 are) in the $ab$ plane (the plane including the $c$ axis)
    The 1--2, 3--4, 4--5, 6--7 transitions are second-order induced by softening of a magnon mode.
    The 3--7, 4--6, and 5--6 transitions are bridged by a high-symmetry line $J_1 = J_2$. The 1--7, 2--5, and 2--6 transitions are bridged by another high-symmetry line $D = 0$.
    The 4--5 boundary is evaluated numerically.
  }
\end{figure}

We first review the case for easy-plane anisotropy, $D < 0$~\cite{Tanaka1989,Kawamura1990,Plumer1991,Zhang1993,Zhitomirsky1995}.
By assuming a three-sublattice structure in the $ab$ plane, we parametrize them as $\mathbf{S}_\mu = (\cos\phi_\mu, \sin\phi_\mu, 0)$ with $\mu \in \{A, B, C\}$. When fixing $\phi_C = 0$ by invoking the U(1) symmetry, the energy density is
\begin{align}
    E(\phi_A,\phi_B) = 3J_1 \left( \cos\phi_A + \cos\phi_B \right) + 3J_2 \cos(\phi_A - \phi_B).
\end{align}
The extremal condition for $E(\phi_A,\phi_B)$ for $J_1 \ne 0$ is $\sin\phi_A + \sin\phi_B = 0$, which can be satisfied by
(i) collinear antiferromagnetic configurations in the honeycomb sublattice $A \cup B$, $(\phi_A,\phi_B,\phi_C)=(\pi, 0, 0)$ or $(0, \pi, 0)$ (hereafter, modulo $2\pi$),
(ii)  collinear states with ferromagnetic configuration in $A \cup B$, $(\phi_A,\phi_B,\phi_C)=(0, 0, 0)$ or $(\pi, \pi, 0)$,
or
(iii) noncollinear states
\begin{align}
     (\phi_A,\phi_B,\phi_C) = \bigl(\pm (\pi/2 + \alpha), \mp(\pi/2 + \alpha),0\bigr),
     \label{eq:easy-plane}
\end{align}
with $\alpha = \sin^{-1}\tfrac{J_1}{2J_2}$, for which $\lvert{J_1/J_2}\rvert < 2$ is required.

\begin{table}
\caption{\label{table:GS-easy-plane}
Symmetry properties of the zero-field ground state phases for $D < 0$ (easy-plane).}
\begin{ruledtabular}
\begin{tabular}{c|cc}
Symmetry &
Phase 1 (collinear) &
Phase 2
 \\[2pt] \hline \\[-0.7em]
    $U(1)$, spin~\footnote{$U(1)$ spin rotation along the $z$ axis.}
&
\None & \None
\\[2.0pt]
%    $\begin{matrix} \text{$Z_2$, spin} \\[-1pt]
%    \text{($S^z \to -S^z$}~\footnote{$\pi$ spin rotation along any axis in the $xy$ plane.}\text{)}
%    \end{matrix}$
%&
%... & ...
%\\[7pt]
    $\begin{matrix} \text{$Z_2$, lattice} \\[-1pt]
    \text{($A \leftrightarrow B$}~\footnote{Mirror symmetry, where the mirror plane contains the $c$ axis, the center of a hexagon, and the midpoint of the hexagon edge.}\text{)}
    \end{matrix}$
&
\Yes & \None
%\\[7pt]
%    $\begin{matrix} \text{$Z_2$, spin-lattice} \\[-1pt]
%    \text{($S^z \to -S^z, A \leftrightarrow B$)}
%    \end{matrix}$
%&
%... & ...
\end{tabular}
\end{ruledtabular}
\end{table}

\begin{table*}
\caption{\label{table:GS-easy-axis}
Symmetry properties of the zero-field ground state phases for $D > 0$ (easy axis).}
\begin{ruledtabular}
\begin{tabular}{c|ccccc}
    &\multicolumn{3}{c}{$0 < \lvert{J_1/J_2}\rvert < 1$}&\multicolumn{2}{c}{$\lvert{J_1/J_2}\rvert > 1$}
    \\[1.5pt]
    Symmetry &
    Phase 3 (collinear)
%    $\begin{matrix}
%    \text{Phase 1 (collinear)} \\[-1pt]
%    \text{up-down-(up/down) mod. $Z_2$}
%    \end{matrix}$
    &
    Phase 4 (noncollinear)
    &
    $\begin{matrix}
    \text{Phase 5 (noncollinear)} \\[-1pt]
    \text{$\mathbf{S}_C \perp c$}
    \end{matrix}$
    &
    $\begin{matrix}
    \text{Phase 6 (noncollinear)} \\[-1pt]
    \text{$\mathbf{S}_C \parallel c$}
    \end{matrix}$
    &
    Phase 7 (collinear)
%    $\begin{matrix}
%    \text{Phase 5 (collinear)} \\[-1pt]
%    \text{up-up-down mod. $Z_2$}
%    \end{matrix}$
\\[2pt] \hline \\[-0.7em]
    $U(1)$, spin~\footnote{$U(1)$ spin rotation along the $z$ axis.}
&
    \Yes & \None & \None & \None & \Yes
\\[2.0pt]
    $\begin{matrix} \text{$Z_2$, spin} \\[-1pt]
    \text{($S^z \to -S^z$}~\footnote{$\pi$ spin rotation along any axis in the $xy$ plane.}\text{)}
    \end{matrix}$
&
    \None & \None & \None & \None & \None
\\[7pt]
    $\begin{matrix} \text{$Z_2$, lattice} \\[-1pt]
    \text{($A \leftrightarrow B$}~\footnote{Mirror symmetry, where the mirror plane contains the $c$ axis, the center of a hexagon, and the midpoint of the hexagon edge.}\text{)}
    \end{matrix}$
&
    \None & \None & \None & \None & \Yes
\\[7pt]
    $\begin{matrix} \text{$Z_2$, spin-lattice} \\[-1pt]
    \text{($S^{x,y} \to -S^{x,y}$}~\footnote{$\pi$ rotation along the $z$ axis.}, \text{$A \leftrightarrow B$)}
    \end{matrix}$
&
    \None & \None & \None & \Yes & \Yes
\\[7pt]
    $\begin{matrix} \text{$Z_2$, spin-lattice} \\[-1pt]
    \text{($S^z \to -S^z$}~\footnote{$\pi$ spin rotation along an axis in the $xy$ plane.}\text{, $A \leftrightarrow B$)}
    \end{matrix}$
&
    \None & \None & \Yes~\footnote{The axis of the $\pi$ spin rotation should coincide with the intersection between the ordering plane and the $xy$ plane, i.e, be parallel to $\mathbf{S}_C$} & \None & \None
\end{tabular}
\end{ruledtabular}
\end{table*}

For $J_2 > 0$ and $\lvert{J_1}\rvert \ge 2J_2$, the ground state is $(\pi, \pi, 0)$ (``Phase 1'' in Fig.~\ref{fig:GS}). For $J_2 > 0$ and  $\lvert{J_1}\rvert < 2J_2$, the ground state is the noncollinear state~\eqref{eq:easy-plane}, which interpolates $(\phi_A,\phi_B,\phi_C)=(0, 0, 0)$ and $(\pi, \pi, 0)$ upon changing $J_1/J_2$ from $J_1 = -2J_2$ to $2J_2$ (``Phase 2''), with an exception of $J_1 = 0$, which has a degenerate ground state manifold due to the decoupling between $A \cup B$ and $C$. The two branches in Eq.~\eqref{eq:easy-plane} may be distinguished by chirality $\chi_z = (\mathbf{S}_A \times \mathbf{S}_B + \mathbf{S}_B \times \mathbf{S}_C + \mathbf{S}_C \times \mathbf{S}_A)\cdot \hat{z}$, corresponding to breaking of mirror symmetry $A \leftrightarrow B$.

The above arguments hold true also for $D = 0$ as far as the relative orientations of the ordered moments are concerned. The only difference is that the direction of the net magnetic moment (the plane of the ordered magnetic moments) for a collinear (noncollinear) state has to be chosen by spontaneous symmetry breaking.

For easy-axis anisotropy, $D > 0$, we first take a perturbative approach assuming $D \ll 1$. We demonstrate below that a $\pi/2$-reorientation transition occurs at $J_1 = J_2$ concerning the noncollinear-coplanar order in Phase 2. Here, instead of taking the direction of the ordering plane as a variational parameter, we consider the rotation of the anisotropy axis by writing $H_D(\mathbf{d}) = -D \sum_i (\mathbf{S}_i \cdot \mathbf{d})^2$ with a variational unit vector $\mathbf{d}$, whilst keeping the unperturbed state in the $ab$ plane [Eq.~\eqref{eq:easy-plane}].
By parametrizing $\mathbf{d}=(\sin\vartheta\cos\varphi,\sin\vartheta\sin\varphi,\cos\vartheta)$ with $0 \le \vartheta \le \pi/2$ and $0 \le \varphi < 2\pi$ ($0 \le \varphi < \pi$ for $\vartheta = \pi/2$), we have
\begin{align}
    H_D(\vartheta,\varphi) = -D \sin^2\vartheta \left( \sin^2(\varphi - \alpha) + \sin^2(\varphi + \alpha) + \cos^2\varphi \right)
\end{align}
for the noncollinar order~\eqref{eq:easy-plane}.
For $J_1 \ne J_2$, the condition of $\partial H_D / \partial \vartheta = \partial H_D / \partial \varphi = 0$ can be satisfied by $\vartheta = 0$, $(\vartheta,\varphi) = (\pi/2,0)$, or $(\vartheta,\varphi) = (\pi/2,\pi/2)$, and the ground state for infinitesimal $D > 0$ is $(\pi/2,\pi/2)$ for $0 < J_1/J_2 < 1$ and $(\pi/2,0)$ for $J_1/J_2 > 1$. In the original coordinate frame, this means that the spin configuration is in a plane that contains the $c$ axis and $\mathbf{S}_C \perp c$ for $0 < J_1/J_2 < 1$ (``Phase 5'') whereas $\mathbf{S}_C \parallel c$ for $J_1/J_2 > 1$ (``Phase 6''). For the easy-axis equilateral triangular lattice ($J_1 = J_2$, $D > 0$), it is well-known that the ground state for $D < (3/2)J$ is the coplanar ``Y'' state in which $\mathbf{S}_\mu$ with chosen spontaneously $\mu \in \{A, B, C\}$ is parallel or antiparallel to the $c$ axis~\cite{Miyashita1986}. In fact, Phase 6 corresponds to a adiabatic deformation of the Y phase for $J_1 = J_2$ while the moment (anti)parallel to the $c$ axis is fixed to sublattice $C$, which is reasonable because the deformed interaction $J_1 > J_2$ is fully compatible with the modulation of the exchange energy in the `Y' state. The spin configuration in Phase 6 continuously changes with $J_1 / J_2$ and becomes collinear at (and above) $J_1/J_2 = 2$ as $(S_A^z,S_B^z,S_C^z)= \pm(1,1,-1)$, which is further stabilized by $D > 0$ (``Phase 7''). The line $D_c = -(3/2)J_1 + 3J_2$ of instability of Phase 7 extends up to $(J_1/J_2, D) = (1, 3/2)$.

Stemming from the line of the decoupled point $J_1 = 0$, there is another collinear phase with four-fold degeneracy, $(S_A^z,S_B^z,S_C^z)=(1,-1,\pm1), (-1,+1,\pm1)$ (``Phase 3''). Stabilized by the easy-axis anisotropy, Phase 3 extends to the $J_1 \ne 0$ region for $D > D'_c =  \frac{3J_2}{2}\bigl(\!\sqrt{3(J_1/J_2)^2 + 1} - 1\bigr)$, bordering on Phase 7 along $J_1 = J_2$, $D > 3/2$ (Fig.~\ref{fig:GS}). Phases 3 and 7 become degenerate at $J_1 = J_2$ and constitute the six-fold degenerate up-up-down (UUD) states.

Phases 3 and 5 touch with each other at $J_1 = D = 0$, though they have quite distinct spin configurations, $\mathbf{S}_C \parallel c$ and $\mathbf{S}_C \perp c$, respectively. This implies there must be either a direct first order transition between them or intermediate phase(s) around $J_1 = D = 0$. Indeed, we find another coplanar noncollinear phase (``Phase 4'') that intervenes the two phases (Fig.~\ref{fig:GS}).
Both 3--4 and 4--5 transitions are second order with the former taking place at $D = D'_c$. Unlike Phase 5, $\mathbf{S}_C$ in Phase 4 is canted off the $ab$ plane to gain anisotropy energy [Fig.~\ref{fig:GS}(a)]. Consequently, Phase 4 breaks the $Z_2$ spin-lattice symmetry that combines mirror transformations $A \leftrightarrow B$ and $S^z \to -S^z$ (more precisely, a $\pi$-rotation along the axis of intersection between the ordering plane and the $ab$ plane), whereas Phase 5 does not. Other symmetry properties for the zero-field ground states discussed above tare summarized in Tables~\ref{table:GS-easy-plane} and \ref{table:GS-easy-axis} for $D < 0$ and $D > 0$, respectively.

\subsection{%
\label{app:subsec:magnetization:c}
Magnetization process for $\mathbf{H} \parallel c$
}
\begin{figure*}
    \begin{center}
    \includegraphics[width=\hsize]{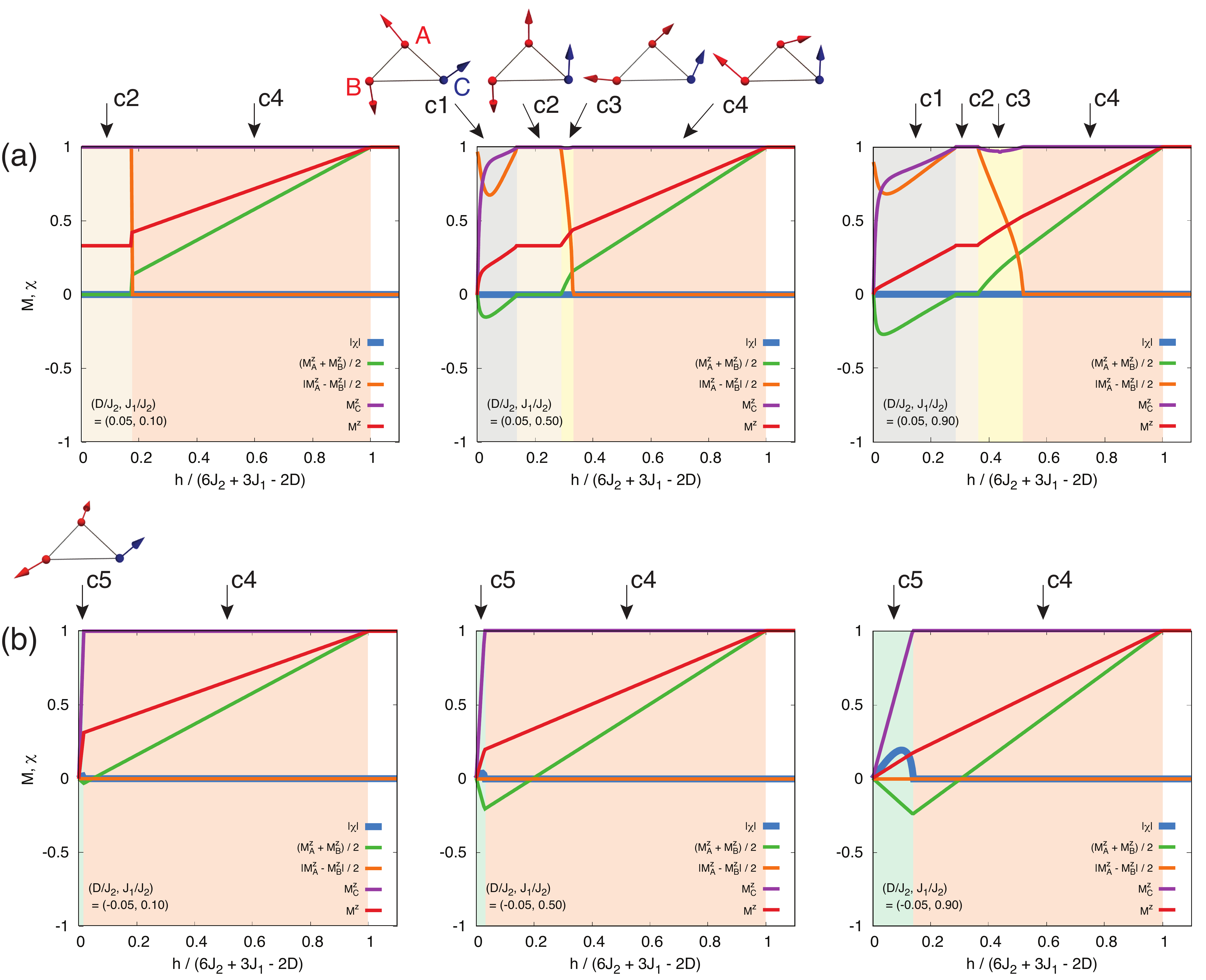}
\end{center}
\caption{
  \label{fig:Mz_small-J}
  Magnetization processes for $\mathbf{h}\parallel c$ for $J_1 < J_2$ with
  (a) $D / J_2 = 0.05$ (small easy-axis) and
  (b) $D / J_2 = -0.05$ (small easy-plane).
  See the text for explanations about the schematic spin configurations.
  }
\end{figure*}

\begin{figure*}
    \begin{center}
    \includegraphics[width=\hsize]{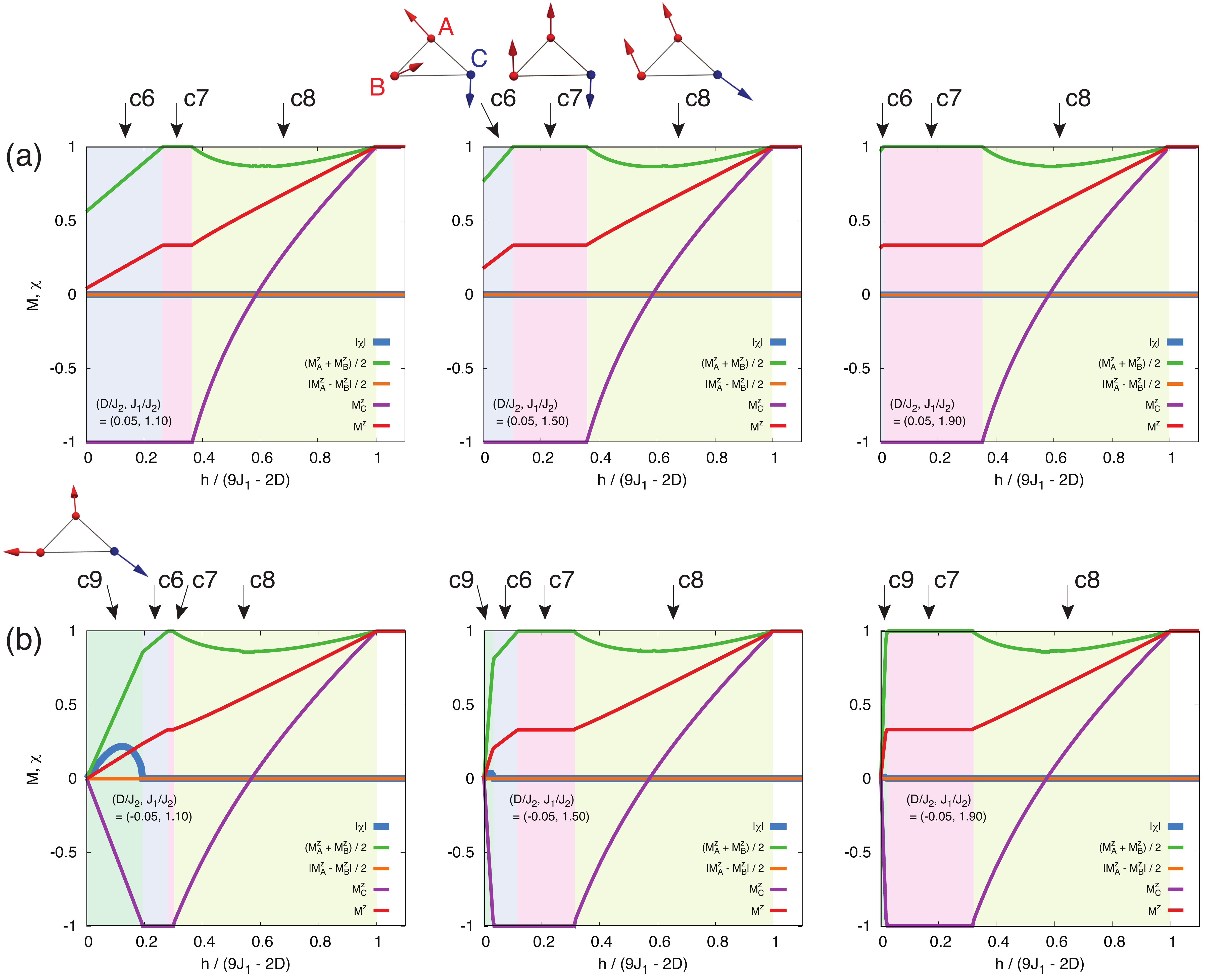}
\end{center}
\caption{
  \label{fig:Mz_large-J}
  Magnetization processes for $\mathbf{h}\parallel c$ for $J_1 > J_2$ with
  (a) $D / J_2 = 0.05$ (small easy-axis) and
  (b) $D / J_2 = -0.05$ (small easy-plane).
  See the text for explanations about the schematic spin configurations.
  }
\end{figure*}

We next discuss the magnetization processes for the field parallel to the $c$ axis.
Figures.~\ref{fig:Mz_small-J} and \ref{fig:Mz_large-J} show representative magnetization curves for small anisotropy $D / J_2 = \pm 0.05$ and selected values of $J_1/J_2$. An immediate observation related to GdInO$_3$ is that a 1/3 magnetization plateau with a UUD configuration appears in a wide region of the parameter space. The UUD state for $J_1 < J_2$ has spin-down in sublattice $A$ or $B$, which we refer to as UUD-$A/B$ [``$c$2'' in Fig.~\ref{fig:Mz_small-J}(a)]. By a stability analysis, we find that the UUD-$A/B$ plateau is stable for any value of $D > 0$ for $J_1 < J_2$. The stability range of this plateau varies as $\propto \sqrt{D}$ for $J_1 < J_2$.
Meanwhile, the UUD state for $J_1 > J_2$ has spin-down in sublattice $C$ (``$c$7'' in Fig.~\ref{fig:Mz_large-J}), which we refer to as UUD-$C$. This state breaks no symmetry for $h^z \ne 0$ and is actually a paramagnetic state. The stability range of the UUD-$C$ state slightly extends to $D<0$ (see below).

Overall, the character of the magnetization process is different between $J_1 > J_2$ and $J_1 < J_2$ and, roughly speaking, between $D > 0$ and $D < 0$. To demonstrate the difference between $J_1 > J_2$ and $J_1 < J_2$, we consider the instability of the fully polarized phase upon approaching the saturation field from above. By introducing complex fields $\mathbf{b} = (b_A\, b_B\, b_C)^\mathrm{T}$ with $\lvert{b_\mu}\rvert \ll 1$, $\mu \in \{A,B,C\}$ to represent spin fluctuations as $\mathbf{S}_\mu = (\mathrm{Re}\,b_\mu, \mathrm{Im}\,b_\mu, \sqrt{1 - \lvert{b_\mu}\rvert^2})$, the mean-field energy density is
$E_\text{FP} \approx E_\text{FP}^0 + \mathbf{b}^\dag \mathcal{D}_\text{FP} \mathbf{b}^{}$
with
\begin{align}
    \mathcal{D}_\text{FP} =
    \begin{pmatrix}
    \frac{h^z-3(J_1 + J_2)}{2} + D &
    \frac{3J_2}{2} &
    \frac{3J_1}{2}
    \\[3pt]
    \frac{3J_2}{2} &
    \frac{h^z-3(J_1 + J_2)}{2} + D &
    \frac{3J_1}{2}
    \\[3pt]
    \frac{3J_1}{2} &
    \frac{3J_1}{2} &
    \frac{h^z}{2} - 3J_1 + D
    \end{pmatrix}
\end{align}
and $E_\text{FP}^0 = 6J_1 + 3J_2 - 3D - 3h^z$. For $J_1 < J_2$, the instability occurs at $h^{z,c}_{J_1 < J_2} = 3 J_1 + 6J_2 - 2D$ and the soft mode is $\mathbf{b}_{J_1 < J_2} \propto(1, -1, 0)$, which is an instability to a coplanar state in a plane including the $c$ axis (``$c$4'' in Fig.~\ref{fig:Mz_small-J}). This spin configuration resembles the Greek letter ``$\Psi$'' when projected in the spin space and, as such, this is also known as the $\Psi$ state in some literature~\cite{Nikuni1995,Starykh2014,Yamamoto2015}.
For $J_1 > J_2$, the instability occurs at $h^{z,c}_{J_1 > J_2} = 9 J_1 - 2D$ and the soft mode is $\mathbf{b}_{J_1 > J_2} \propto (1,1,-2)$. This leads to the so-called ``V'' state (``$c$8'' in Fig.~\ref{fig:Mz_large-J})~\cite{Nikuni1995}, which is coplanar in the plane including the $c$ axis. These results hold true for both $D > 0$ and $D < 0$.

In the following, we quickly go through further details about the magnetization processes for $J_1 < J_2$ and $J_1 > J_2$ with both signs of $D$. For $J_1 < J_2$ and $D < 0$ [Fig.~\ref{fig:Mz_small-J}(b)], the low-field regime realizes a noncoplanar phase resulting from canting of the zero-field state in Phase 2 out of the $ab$ plane [``c5'' in Fig.~\ref{fig:Mz_small-J}(b)]. Because the antiferromagnetic exchange for sublattice $C$ is smaller than the other two for $J_1 < J_2$, $M_C^z$ in this phase increases rapidly as $h$ increased. The scalar chirality $\chi = \mathbf{S}_A \cdot \mathbf{S}_B \times \mathbf{S}_C$ also varies with the magnetic field and vanishes at the same magnetic field where $M_C^z$ is saturated. This is a second-order noncoplanar-coplanar transition into the $\Psi$ state discussed above.

For $J_1 < J_2$ and $D > 0$ [Fig.~\ref{fig:Mz_small-J}(a)], we have seen that the UUD-$A/B$ state and the $\Psi$ state are realized in the intermediate and high field regimes, respectively. Interestingly, for a symmetry reason, they cannot be smoothly deformed from one to the other. For example, $\lvert{M^z_A - M^z_B}\rvert$ takes its maximum value in the UUD-$A/B$ ($M^z_\mu = \langle{S^z_\mu}\rangle$, $\mu \in \{A,B,C\}$) while it is zero in the $\Psi$ state [Fig.~\ref{fig:Mz_small-J}(a)]. Therefore, there must be a direct first order transition, otherwise some intermediate phase(s). For relatively large values of $J_1$, e.g., $J_1 / J_2 = 0.5, 0.9$ for $D / J_2 = 0.05$ shown in Fig.~\ref{fig:Mz_small-J}(a), there is an intermediate phase that emerges via softening of the UUD-$A/B$ state at its high-field edge of the plateau [``$c$3'' in Fig.~\ref{fig:Mz_small-J}(a)], which then undergoes a transition into the $\Psi$ state at a higher magnetic field. Meanwhile, the softening of the UUD-$A/B$ state at the low-field edge of the plateau yields another coplanar state [``c1'' in Fig.~\ref{fig:Mz_small-J}(a)], which is essentially the same state as the one in Phase 4 in zero field.
For smaller values of $J_1$, e.g., $J_1 / J_2 = 0.1$ for $D / J_2 = 0.05$, a direct first-order transition between the UUD-$A/B$ and $\Psi$ states is realized [Fig.~\ref{fig:Mz_small-J}(a)].

\begin{figure*}
    \begin{center}
    \includegraphics[width=\hsize]{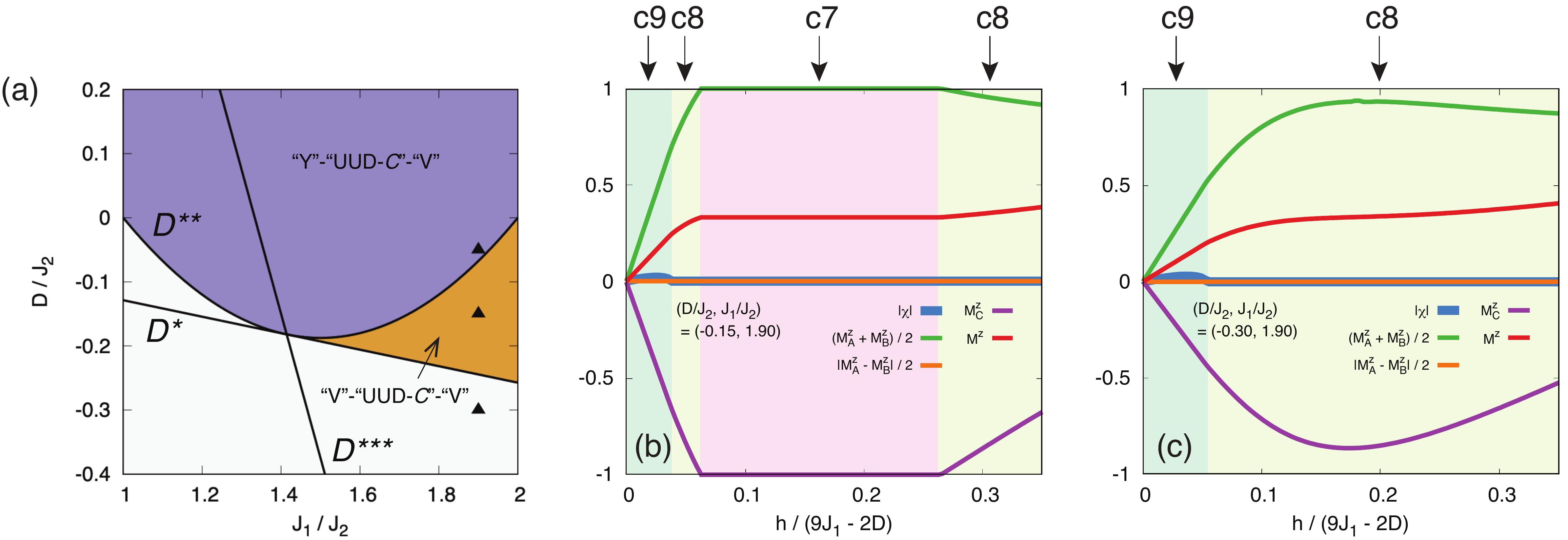}
\end{center}
\caption{
  \label{fig:Mz_large-J:UUD-C}
  (a) Parameter regimes for realizing two distinct types of the (in)stability of the UUD-$C$ state (see the text).
  The triangles indicate the parameter set presented in Fig.~\ref{fig:Mz_large-J}(b) and panels (b) and (c) in this figure.
  (b) Low-field magnetization processes for $\mathbf{h}\parallel c$ for
  $(D/J_2, J_1 / J_2) = (-0.15,1.9)$
  and
  (c)
  the same for $(D/J_2, J_1 / J_2) = (-0.30,1.9)$.
  See Fig.~\ref{fig:Mz_large-J} for the spin state indices.
  }
\end{figure*}

For $J_1 > J_2$ and $D > 0$ [Fig.~\ref{fig:Mz_large-J}(a)], the magnetization process is qualitatively similar to the well-known case of the equilateral triangular-lattice Heisenberg model with easy-axis anisotropy~\cite{Miyashita1986}; the Y state in the low-field regime, the UUD (UUD-$C$) state in the middle, and the V state in the high-field regime [``$c$6'', ``$c$7'', and ``$c$8'', respectively, in Fig.~\ref{fig:Mz_large-J}(a)]. The only caveat is that the $Z_3$ symmetry is broken explicitly for $J_1 \ne J_2$. As discussed earlier about the UUD-$C$ state, this qualitative similarity can be understood via the compatibility between the modulation in the coupling constants and the modulation in the exchange energy in each state.

Finally, for $J_1 > J_2$ and $D < 0$ [Fig.~\ref{fig:Mz_large-J}(b)], we discuss some details about the stability of the UUD-$C$ state (1/3 magnetization plateau), because it is not very common to have a plateau state in a classical system with easy-plane anisotropy.
For this purpose, we write $\mathbf{S}_\mu = (\mathrm{Re}\,b'_\mu, \mathrm{Im}\,b'_\mu, \sigma_\mu \sqrt{1 - \lvert{b'_\mu}\rvert^2})$
with
$\sigma_A = \sigma_B = -\sigma_C = 1$
and
$\lvert{b'_\mu}\rvert \ll 1$, $\mu \in \{A,B,C\}$. The mean-field energy density is
$E_\text{UUD-$C$} \approx E_\text{UUD-$C$}^0 + \mathbf{b}'{^\dag} \mathcal{D}_\text{UUD-$C$} \mathbf{b}'$ with $\mathbf{b}' = (b'_A\, b'_B\, b'_C)^\mathrm{T}$,

\begin{align}
    \mathcal{D}_\text{UUD-$C$} =
    \begin{pmatrix}
    \frac{h^z+3(J_1 - J_2)}{2} + D &
    \frac{3J_2}{2} &
    \frac{3J_1}{2}
    \\[3pt]
    \frac{3J_2}{2} &
    \frac{h^z+3(J_1 - J_2)}{2} + D &
    \frac{3J_1}{2}
    \\[3pt]
    \frac{3J_1}{2} &
    \frac{3J_1}{2} &
    -\frac{h^z}{2} + 3J_1 + D
    \end{pmatrix},
\end{align}
and $E_\text{UUD-$C$}^0 = -6J_1 + 3J_2 - 3D - 3h^z$.
The eigenvalues $\omega'_{1}$--$\omega'_{3}$ of $\mathcal{D}_\text{UUD-$C$}$ (corresponding to the spin wave frequencies at momentum $\mathbf{q} = 0$) are
\begin{align}
\omega'_{1} &= \frac{1}{4} \left(9J_1 - \sqrt{4{h^z}^2 - 12 J_1 h^z + 81 J_1^2}\right) + D,
\notag\\
\omega'_{2} &= \frac{1}{4} \left(9J_1 + \sqrt{4{h^z}^2 - 12 J_1 h^z + 81 J_1^2}\right) + D \ge \omega'_{1},
\notag\\
\omega'_3 &= \frac{3}{2}J_1 - 3J_2 + \frac{h^z}{2} + D.
\end{align}
The condition for $ \omega'_1 > 0$ is
$D > D^{\ast} = -\frac{9 - 6\sqrt{2}}{4}J_1$
and
$h^z_{1-} < h^z < h^z_{1+}$ with $h^z_{1\pm} = \frac{3}{2}J_1 \pm \frac{1}{2}\sqrt{16D^2 + 72 J_1 D + 9 J_1^2}$, whereas the condition for $ \omega'_3 > 0$ is simply $h^z > h^z_3 = -3 J_1 + 6 J_2 - 2D$.
There are two qualitatively distinct scenarios for the stability. If $h^z_{1-} < h^z_3 < h^z_{1+}$, the low- (high-)field instability is caused by the softening of $\omega'_3$ ($\omega'_1$) mode, leading to the Y (V) state, similar to the easy-axis case. The condition for this scenario is $D > D^{\ast\ast} = [\frac{3}{4}(J_1/J_2)^2 - \frac{9}{4}(J_1/J_2) + \frac{3}{2}]J_2$, where we note $D^{\ast\ast} \ge D^{\ast}$ [see Fig.~\ref{fig:Mz_large-J:UUD-C}(a)]. In Fig.~\ref{fig:Mz_large-J}(b), we demonstrate this scenario for $(D/J_2, J_1 / J_2) = (-0.05,1.1)$ and $(-0.05,1.5)$.
If $h^z_3 < h^z_{1-}$, on the other hand, the instabilities at both edges of the plateau are caused by softening of $\omega'_1$ mode. This means that the V state is realized on both sides of the UUD-$C$ phase in this case. This scenario requires $D^{\ast} < D < D^{\ast\ast}$ and $D > D^{\ast\ast\ast} = -(9/4)J_1 + 3J_2$, satisfying both of which is possible only for $J_1 > \!\!\sqrt{2}J_2$ [Fig.~\ref{fig:Mz_large-J:UUD-C}(a)]. We show a magnetization process in the low-field regime for $(D/J_2, J_1 / J_2) = (-0.15,1.9)$ in Fig.~\ref{fig:Mz_large-J:UUD-C}(b) to demonstrate this scenario. By further increasing the easy-plane anisotropy, the plateau becomes narrow and eventually disappears, as demonstrated in Fig.~\ref{fig:Mz_large-J:UUD-C}(c) for $(D/J_2, J_1 / J_2) = (-0.30,1.9)$.
Lastly, to conclude our discussion on the case with $J_1 > J_2$ and $D < 0$, we mention that a noncoplanar state resulting from canting of the zero-field state out of the $ab$ plane is realized in the lowest field regime [``c9'' in Fig.~\ref{fig:Mz_large-J}(b)], similar to the case with $J_1 < J_2$ and $D < 0$. The noncoplanar-coplanar transition into either the Y state or the V state is of the second order.

\subsection{%
\label{app:subsec:magnetization:ab}
Magnetization process for $\mathbf{H} \perp c$
}
\begin{figure*}
    \begin{center}
    \includegraphics[width=\hsize]{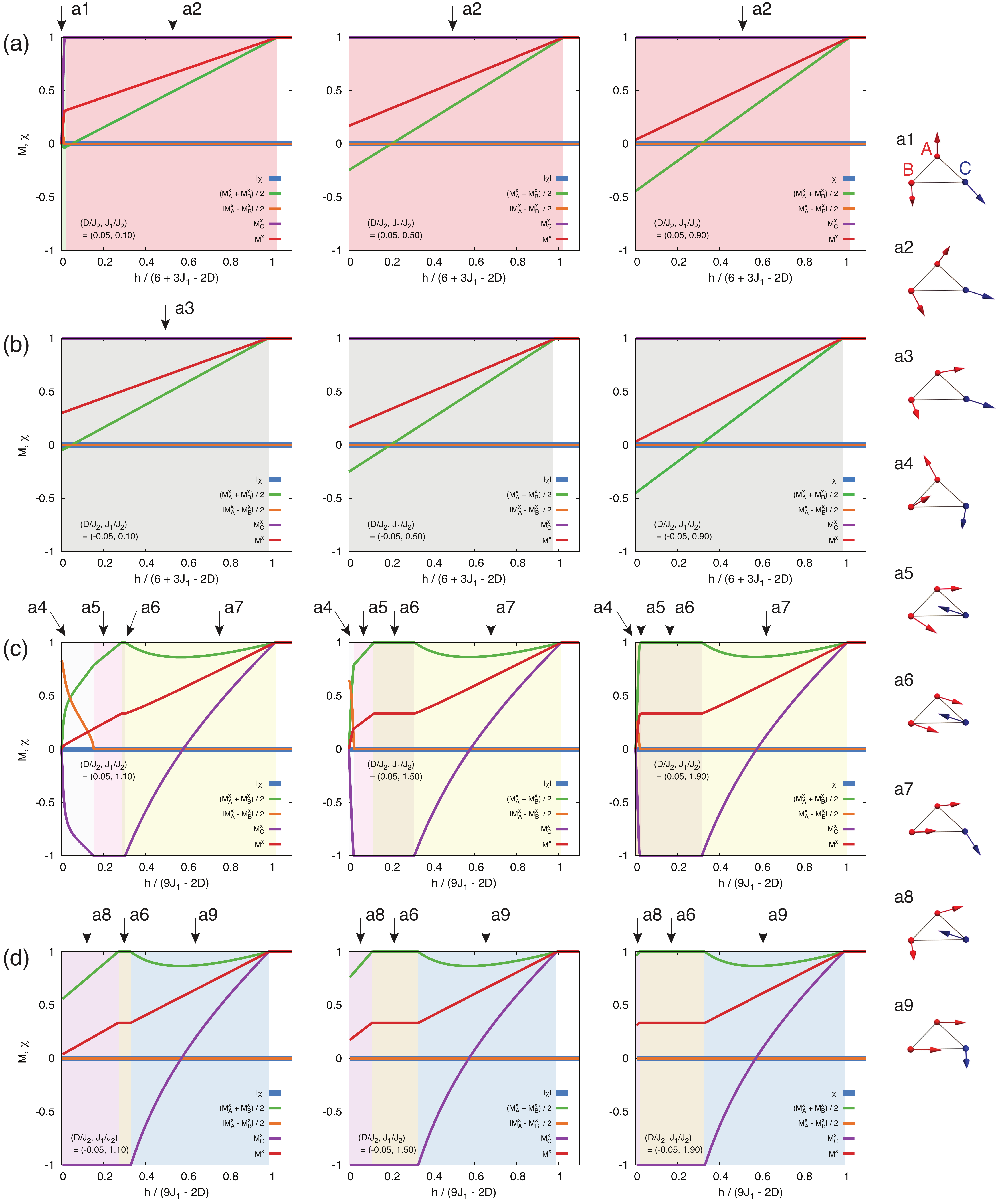}
\end{center}
\caption{
  \label{fig:Mx}
  Magnetization processes for $\mathbf{h} \parallel a$ (or any direction of $\mathbf{h}\perp c$)
  for $J_1 > J_2$ with
  (a) $D / J_2 = 0.05$ (small easy-axis) and $J_1 < J_2$,
  (b) $D / J_2 = -0.05$ (small easy-plane) and $J_1 < J_2$,
  (c) $D / J_2 = 0.05$ and $J_1 > J_2$,
  and
  (d) $D / J_2 = -0.05$ and $J_1 > J_2$.
    See the text for explanations about the schematic spin configurations.
    }
\end{figure*}

Figure~\ref{fig:Mx} shows the magnetization process of the spin model~\eqref{eq:H} in the in-plane magnetic field. In the following, brief explanations about the realized spin configurations are in order. The state ``$a1$'' is a coplanar state, not symmetric under exchanging $A \leftrightarrow B$, in the plane subtended by the direction of $\mathbf{h}$ (i.e., the $a$ axis) and the $c$ axis. In this phase, $M^x_C$ increases fast to reach the saturation as increasing $h^x$. The states ``$a2$'' and ``$a3$'' are $\Psi$ states in the $ac$ and $ab$ planes, respectively. The state ``$a4$'' is a coplanar state in the $ac$ plane and is not symmetric under $A \leftrightarrow B$, which in fact has the same symmetry property as $a1$. The states ``$a5$'' and ``$a6$'' are a Y state in the $ac$ plane and an UUD state parallel to the $a$ axis, respectively, in both of which the magnetic moment in the sublattice $C$ is antiparallel to the field direction. The state ``$a7$'' is a V state in the $ac$ plane.
The states ``$a8$'' and ``$a9$'' are a Y state and a V state both in the $ab$ plane.

\begin{acknowledgments}
We express our sincere thanks to Xianghong Ruan for the delicate graphics production.
Y.K.~acknowledges the support by the NSFC (No.~11950410507, No.~12074246, and No.~U2032213) and MOST (No.~2016YFA0300500 and No.~2016YFA0300501) research programs.
Y.M. Cao acknowledges the support by the National Natural Science Foundation of China (NSFC, No. 51862032) and the Project for Applied Basic Research Programs of Yunnan Province (No. 2018FB010).  K. Xu
acknowledges the support by the National Natural Science Foundation of China (NSFC, No. 51861032), the Project for Applied Basic Research Programs of Yunnan Province (No. 2018FB010) and the Yunnan Local Colleges Applied Basic Research Projects of Yunnan Province (No. 2018FH001-001).
G. Hong acknowledges the support by the research fund of University of Macau, Macau SAR (File no. MYRG2018-00079-IAPME, MYRG2019-00115-IAPME), and the Science and Technology Development Fund, Macau SAR (File no. 081/2017/A2, 0059/2018/A2, 009/2017/AMJ).
\end{acknowledgments}

\bibliographystyle{apsrev4-2}
\bibliography{ref}

\end{document}